\documentclass[11pt]{article}
\usepackage{url,epsfig}
\usepackage{geometry} 
\usepackage{sectsty} 
\usepackage[normalem]{ulem} 
\sectionfont{\sffamily\bfseries\upshape\large}
\subsectionfont{\sffamily\bfseries\upshape\normalsize} 
\subsubsectionfont{\sffamily\bfseries\upshape\normalsize} 
\makeatletter

  \renewcommand\@seccntformat[1]{\csname the#1\endcsname.\quad}

\makeatother\renewcommand{\bibitem}{\vskip 2pt\par\hangindent\parindent\hskip-\parindent}

\makeatletter
\def\@maketitle{%
  \begin{center}%
  \let \footnote \thanks
    {\large \@title \par}%
    {\normalsize
      \begin{tabular}[t]{c}%
        \@author
      \end{tabular}\par}%
    {\small \@date}%
  \end{center}%
}
\makeatother

\newenvironment{example}[1]{\begin{quote}{\bf Example.  #1}\\}{\end{quote}}

\newcommand{\newc}{\newcommand}
\newc{\nnb}{\nonumber}
\newc{\beqann}{\begin{eqnarray*}}
\newc{\beqa}{\begin{eqnarray}}
\newc{\beqnn}{\begin{displaymath}}
\newc{\beq}{\begin{equation}}
\newc{\bex}{\begin{example}}
\newc{\eex}{\end{example}}
\newc{\eeqann}{\end{eqnarray*}}
\newc{\eeqa}{\end{eqnarray}}
\newc{\eeqnn}{\end{displaymath}}
\newc{\eeq}{\end{equation}}
\newc{\E}{\mbox{E}}
\newc{\Invc}{\mbox{Inv-$\chi^2$}}       
\newc{\Nor}{\mbox{N}}
\newc{\Var}{\mbox{var}}
\newc{\btab}{\begin{tabular}}
\newc{\etab}{\end{tabular}}
\newc{\rep}{\rm rep}

\title{\bf Understanding predictive information criteria for Bayesian models\footnote{To appear in {\em Statistics and Computing}.  We thank two reviewers for helpful comments and the National Science Foundation, Institute of Education Sciences, and Academy of Finland (grant 218248) for partial support of this research.}\vspace{.1in}}
\author{Andrew Gelman\footnote{Department of Statistics, Columbia University, New York, N.Y.}, Jessica Hwang\footnote{Department of Statistics, Harvard University, Cambridge, Mass.}, and Aki Vehtari\footnote{Department of Biomedical Engineering and Computational Science, Aalto University, Espoo, Finland.}}
\date{22 July 2013}
\begin{document}
\maketitle
\thispagestyle{empty}

\begin{abstract}
We review the Akaike, deviance, and Watanabe-Akaike information criteria from a Bayesian perspective, where the goal is to estimate expected out-of-sample-prediction error using a bias-corrected adjustment of within-sample error.  We focus on the choices involved in setting up these measures, and we compare them in three simple examples, one theoretical and two applied.  The contribution of this review is to put all these information criteria into a Bayesian predictive context and to better understand, through small examples, how these methods can apply in practice.

Keywords:  AIC, DIC, WAIC, cross-validation, prediction, Bayes
\end{abstract}

\section{Introduction}

Bayesian models can be evaluated and compared in several ways.  Most simply, any model or set of models can be taken as an exhaustive set, in which case all inference is summarized by the posterior distribution.  The fit of model to data can be assessed using posterior predictive checks (Rubin, 1984), prior predictive checks (when evaluating potential replications involving new parameter values), or, more generally, mixed checks for hierarchical models (Gelman, Meng, and Stern, 2006).  When several candidate models are available, they can be compared and averaged using Bayes factors (which is equivalent to embedding them in a larger discrete model) or some more practical approximate procedure (Hoeting et al., 1999) or continuous model expansion (Draper, 1999).

In other settings, however, we seek not to check models but to compare them and explore directions for improvement.  Even if all of the models being considered have mismatches with the data, it can be informative to evaluate their predictive accuracy, compare them, and consider where to go next.  The challenge then is to estimate predictive model accuracy, correcting for the bias inherent in evaluating a model's predictions of the data that were used to fit it.

A natural way to estimate out-of-sample prediction error is cross-validation (see Vehtari and Lampinen, 2002, for a Bayesian perspective), but researchers have always sought alternative measures, as cross-validation requires repeated model fits and can run into trouble with sparse data.  For practical reasons alone, there remains a place for simple bias corrections such as AIC (Akaike, 1973), DIC (Spiegelhalter et al., 2002, van der Linde, 2005), and, more recently, WAIC (Watanabe, 2010), and all these can be viewed as approximations to different versions of cross-validation (Stone, 1977).

At the present time, DIC appears to be the predictive measure of choice in Bayesian applications, in part because of its incorporation in the popular BUGS package (Spiegelhalter et al., 1994, 2003).  Various difficulties have been noted with DIC (see Celeux et al., 2006, Plummer, 2008, and much of the discussion of Spiegelhalter et al., 2002) but there has been no consensus on an alternative.

One difficulty is that all the proposed measures are attempting to perform what is, in general, an impossible task:  to obtain an unbiased (or approximately unbiased) and accurate measure of out-of-sample prediction error that will be valid over a general class of models and that requires minimal computation beyond that needed to fit the model in the first place.  When framed this way, it should be no surprise to learn that no such ideal method exists.  But we fear that the lack of this panacea has impeded practical advances, in that applied users are left with a bewildering array of choices.

The purpose of the present article is to explore AIC, DIC, and WAIC from a Bayesian perspective in some simple examples.  Much has been written on all these methods in both theory and practice, and we do not attempt anything like a comprehensive review (for that, see Vehtari and Ojanen, 2012).  Our unique contribution here is to view all these methods from the standpoint of Bayesian practice, with the goal of understanding certain tools that are used to understand models.  We work with three simple (but, it turns out, hardly trivial) examples to develop our intuition about these measures in settings that we understand.  We do not attempt to derive the measures from first principles; rather, we rely on the existing literature where these methods have been developed and studied.

In some ways, our paper is similar to the review article by Gelfand and Dey (1994), except that they were focused on model choice whereas our goal is more immediately to estimate predictive accuracy for the goal of model comparison.  As we shall discuss in the context of an example, given the choice between two particular models, we might prefer the one with higher expected predictive error; nonetheless we see predictive accuracy as one of the criteria that can be used to evaluate, understand, and compare models.

\section{Log predictive density as a measure of model accuracy}

One way to evaluate a model is through the accuracy of its predictions.  Sometimes we care about this accuracy for its own sake, as when evaluating a forecast.  In other settings, predictive accuracy is valued not for its own sake but rather for comparing different models.  We begin by considering different ways of defining the accuracy or error of a model's predictions, then discuss methods for estimating predictive accuracy or error from data. 

\subsection{Measures of predictive accuracy}

Consider data $y_1,\dots,y_n$, modeled as independent given parameters $\theta$; thus $p(y|\theta)=\prod_{i=1}^np(y_i|\theta)$.  With regression, one would work with $p(y|\theta,x)=\prod_{i=1}^np(y_i|\theta,x_i)$.  In our notation here we suppress any dependence on $x$.

Preferably, the measure of predictive accuracy is specifically tailored for the application at hand, and it measures as correctly as possible the benefit (or cost) of predicting future data with the model. Often explicit benefit or cost information is not available and the predictive performance of a model is assessed by generic scoring functions and rules.

Measures of predictive accuracy for point prediction are called scoring functions. A good review of the most common scoring functions is presented by Gneiting (2011), who also discusses the desirable properties for scoring functions in prediction problems.  We use the squared error as an example scoring function for point prediction, because the squared error and its derivatives seem to be the most common scoring functions in predictive literature (Gneiting, 2011).

Measures of predictive accuracy for probabilistic prediction are called scoring rules.  Examples include the quadratic, logarithmic, and zero-one scores, whose properties are reviewed by Gneiting and Raftery (2007). Bernardo and Smith (1994) argue that suitable scoring rules for prediction are proper and local: propriety of the scoring rule motivates the decision maker to report his or her beliefs honestly, and for local scoring rules predictions are judged only on the plausibility they assign to the event that was actually observed, not on predictions of other events.
The logarithmic score is the unique (up to an affine transformation) local and proper scoring rule (Bernardo, 1979), and appears to be the most commonly used scoring rule in model selection.

\paragraph{Mean squared error.}
A model's fit to new data can be summarized numerically by mean squared error,
$\frac{1}{n}\sum_{i=1}^n(y_i-\E(y_i|\theta))^2$, or a weighted version such as
$\frac{1}{n}\sum_{i=1}^n(y_i-\E(y_i|\theta))^2/\Var(y_i|\theta)$.  These measures have the advantage of being easy to compute and, more importantly, to interpret, but the disadvantage of being less appropriate for models that are far from the normal distribution.

\paragraph{Log predictive density or log-likelihood.}

A more general summary of predictive fit is the log predictive density,  $\log p(y|\theta)$,
which is proportional to the mean squared error if the model is normal with constant variance.  The log predictive density is also sometimes called the log-likelihood. The log predictive density has an important role in statistical model comparison because of its connection to the Kullback-Leibler information measure (see Burnham and Anderson, 2002, and Robert, 1996).
In the limit of large sample sizes, the model with the lowest Kullback-Leibler information---and thus, the highest expected log predictive density---will have the highest posterior probability.  Thus, it seems reasonable to use expected log predictive density as a measure of overall model fit.

Given that we are working with the log predictive density, the question may arise:  why not use the log posterior?  Why only use the data model and not the prior density in this calculation?  The answer is that we are interested here in summarizing the fit of model {\em to data}, and for this purpose the prior is relevant in estimating the parameters but not in assessing a model's accuracy.

We are not saying that the prior cannot be used in assessing a model's fit to data; rather we say that the prior density is not relevant in computing {\em predictive} accuracy. Predictive accuracy is not the only concern when evaluating a model, and even within the bailiwick of predictive accuracy, the prior is relevant in that it affects inferences about $\theta$ and thus affects any calculations involving $p(y|\theta)$.  In a sparse-data setting, a poor choice of prior distribution can lead to weak inferences and poor predictions.

\subsection{Log predictive density asymptotically, or for normal linear models}

Under standard conditions, the posterior distribution, $p(\theta|y)$, approaches a normal distribution in the limit of increasing sample size (see, e.g., DeGroot, 1970).  In this asymptotic limit, the posterior is dominated by the likelihood---the prior contributes only one factor, while the likelihood contributes $n$ factors, one for each data point---and so the likelihood function also approaches the same normal distribution.

As sample size $n\!\rightarrow\!\infty$, we can label the limiting posterior distribution as $\theta|y\rightarrow \Nor\left(\theta_0,V_0/n\right)$.  In this limit the log predictive density is
\beqnn
\log p(y|\theta)=c(y)-\frac{1}{2}\left(k\log(2\pi) + \log |V_0/n| +(\theta-\theta_0)^T(V_0/n)^{-1}(\theta-\theta_0)\right),
\eeqnn
where $c(y)$ is a constant that only depends on the data $y$ and the model class but not on the parameters $\theta$. 

The limiting multivariate normal distribution for $\theta$ induces a posterior distribution for the log predictive density that ends up being a constant (equal to
$c(y)-\frac{1}{2}\left(k\log(2\pi) + \log |V_0/n|\right)$) minus $\frac{1}{2}$ times a $\chi^2_k$ random variable, where $k$ is the dimension of $\theta$, that is, the number of parameters in the model.  The maximum of this distribution of the log predictive density is attained when $\theta$ equals the maximum likelihood estimate (of course), and its posterior mean is at a value $\frac{k}{2}$ lower.  For actual posterior distributions, this asymptotic result is only an approximation, but it will be useful as a benchmark for interpreting the log predictive density as a measure of fit.

With singular models (e.g. mixture models and overparameterized complex models more generally) a set of different parameters can map to a single data model, the Fisher information matrix is not positive definite, plug-in estimates are not representative of the posterior, and the distribution of the deviance does not converge to a $\chi^2$ distribution. The asymptotic behavior of such models can be analyzed using singular learning theory (Watanabe, 2009, 2010).

\subsection{Predictive accuracy for a single data point}\label{expectedaccuracy}

The ideal measure of a model's fit would be its out-of-sample predictive performance for new data produced from the true data-generating process.  We label $f$ as the true model, $y$ as the observed data (thus, a single realization of the dataset $y$ from the distribution $f(y)$), and $\tilde{y}$ as future data or alternative datasets that could have been seen. The out-of-sample predictive fit for a new data point $\tilde{y}_i$ using logarithmic score is then,
\beqnn
\log p_{\rm post}(\tilde{y}_i) = \log \E_{\rm post}(p(\tilde{y}_i|\theta)) = \log \int \!p(\tilde{y}_i|\theta)p_{\rm post}(\theta)d\theta.
\eeqnn
In the above expression, $p_{\rm post}(\tilde{y}_i)$ is the predictive density for $\tilde{y}_i$ induced by the posterior distribution $p_{\rm post}(\theta)$.  We have introduced the notation $p_{\rm post}$ here to represent the posterior distribution because our expressions will soon become more complicated and it will be convenient to avoid explicitly showing the conditioning of our inferences on the observed data $y$.  More generally, we use $p_{\rm post}$ and $\E_{\rm post}$ to denote any probability or expectation that averages over the posterior distribution of $\theta$.

We must then take one further step.  The future data $\tilde{y}_i$ are themselves unknown and thus we define the expected out-of-sample log predictive density,
\beqa\nnb
\mbox{elpd} &=& \mbox{expected log predictive density for a new data point}\\
\label{elppd} &=& 
\E_f (\log p_{\rm post}(\tilde{y}_i)) = \int (\log p_{\rm post}(\tilde{y}_i)) f(\tilde{y}_i)d\tilde{y}.
\eeqa
In the machine learning literature this is often called the mean log predictive density.
In any application, we would have some $p_{\rm post}$ but we do not in general know the data distribution $f$.  A natural way to estimate the expected out-of-sample log predictive density would be to plug in an estimate for $f$, but this will tend to imply too good a fit, as we discuss in Section \ref{section.waic}.  For now we consider the estimation of predictive accuracy in a Bayesian context.

To keep comparability with the given dataset, one can define a measure of predictive accuracy for the $n$ data points taken one at a time:
\beqa\nnb
\mbox{elppd} &=& \mbox{expected log pointwise predictive density for a new dataset}\\
\label{elppd_2} &=& \sum_{i=1}^n \E_f (\log p_{\rm post}(\tilde{y}_i)),
\eeqa
which must be defined based on some agreed-upon division of the data $y$ into individual data points $y_i$.
The advantage of using a pointwise measure, rather than working with the joint posterior predictive distribution, $p_{\rm post}(\tilde{y})$ is in the connection of the pointwise calculation to cross-validation, which allows some fairly general approaches to approximation of out-of-sample fit using available data.

It is sometimes useful to consider predictive accuracy given a point estimate $\hat{\theta}(y)$, thus,
\beq\label{elpepd}
\mbox{expected log predictive density, given $\hat{\theta}$: }  \ E_f (\log p(\tilde{y}|\hat{\theta})).
\eeq
For models with independent data given parameters, there is no difference between joint or pointwise prediction given a point estimate, as $p(\tilde{y}|\hat{\theta}) = \prod_{i=1}^n p(\tilde{y}_i|\hat{\theta})$.

\subsection{Evaluating predictive accuracy for a fitted model}

In practice the parameter $\theta$ is not known, so we cannot know the log predictive density $\log p(y|\theta)$.  For the reasons discussed above we would like to work with the posterior distribution, $p_{\rm post}(\theta)=p(\theta|y)$, and summarize the predictive accuracy of the fitted model to data by
\begin{eqnarray}\label{lppd}
\nonumber \mbox{lppd} &=& \mbox{log pointwise predictive density}\\
&=&  \log \prod_{i=1}^n p_{\rm post}(y_i)=\sum_{i=1}^n \log\!\int \!p(y_i|\theta)p_{\rm post}(\theta)d\theta.
\end{eqnarray}
To compute this predictive density in practice, we can evaluate the expectation using draws from $p_{\rm post}(\theta)$, the usual posterior simulations, which we label $\theta^s$, $s=1,\dots,S$:
\beqa\nnb
\mbox{computed lppd} &=&\mbox{computed log pointwise predictive density} \\
&=& \sum_{i=1}^n \log \left(\frac{1}{S}\sum_{s=1}^S
p(y_i|\theta^s)\right). \label{butgut3}
\eeqa
We typically assume that the number of simulation draws $S$ is large enough to fully capture the posterior distribution; thus we shall refer to the theoretical value (\ref{lppd}) and the computation (\ref{butgut3}) interchangeably as the log pointwise predictive density or lppd of the data.

As we shall discuss in Section \ref{section.waic}, the lppd of observed data $y$ is an underestimate of the elppd for future data (\ref{elppd_2}).  Hence the plan is to like to start with (\ref{butgut3}) and then apply some sort of bias correction to get a reasonable estimate of (\ref{elppd_2}).

\subsection{Choices in defining the likelihood and predictive quantities}
As is well known in hierarchical modeling (see, e.g., Spiegelhalter et al., 2002, Gelman et al., 2003),
the line separating prior distribution from likelihood is somewhat arbitrary and is related to the question of what aspects of the data will be changed in hypothetical replications.  In a hierarchical model with direct parameters $\alpha_1,\ldots,\alpha_J$ and hyperparameters $\phi$, factored as $p(\alpha,\phi|y)\propto p(\phi)\prod_{j=1}^J p(\alpha_j|\phi)p(y_j|\alpha_j)$, we can imagine replicating new data in existing groups (with the `likelihood' being proportional to $p(y|\alpha_j)$) or new data in new groups (a new $\alpha_{J+1}$ is drawn, and the `likelihood' is proportional to $p(y|\phi)=\int\! p(\alpha_{J+1}|\phi)p(y|\alpha_{J+1})d\alpha_{J+1}$).  In either case we can easily compute the posterior predictive density of the observed data $y$:
\begin{itemize}
\item When predicting $\tilde{y}|\alpha_j$ (that is, new data from existing groups), we compute $p(y|\alpha_j^s)$ for each posterior simulation $\alpha_j^s$ and then take the average, as in (\ref{butgut3}).
\item When predicting $\tilde{y}|\alpha_{J+1}$ (that is, new data from a new group), we sample $\alpha_{J+1}^s$ from $p(\alpha_{J+1}|\phi^s)$ to compute $p(y|\alpha_{J+1}^s)$.
\end{itemize}
Similarly, in a mixture model, we can consider replications conditioning on the mixture indicators, or replications in which the mixture indicators are redrawn as well.

Similar choices arise even in the simplest experiments.  For example, in the model $y_1,\dots,y_n\sim \Nor(\mu,\sigma^2)$, we have the option of assuming the sample size is fixed by design (that is, leaving $n$ unmodeled) or treating it as a random variable and allowing a new $\tilde{n}$ in a hypothetical replication.

We are not bothered by the nonuniqueness of the predictive distribution.  Just as with posterior predictive checks (Rubin, 1984), different distributions correspond to different potential uses of a posterior inference. Given some particular data, a model might predict new data accurately in some scenarios but not in others.

Vehtari and Ojanen (2012) discus different prediction scenarios where the future explanatory variable $\tilde{x}$ is assumed to be random, unknown, fixed, shifted, deterministic, or constrained in some way. Here we consider only scenarios with no $x$, $p(\tilde{x})$ is equal to $p(x)$, or $\tilde{x}$ is equal to $x$. Variations of cross-validation and hold-out methods can be used for more complex scenarios. For example, for time series with unknown finite range dependencies, $h$-block cross-validation (Burman et al., 1994) can be used. Similar variations of information criteria have not been proposed. Regular cross-validation and information criteria can be used for time series in case of stationary Markov process and squared error or a scoring function or rule which is well approximated by a quadratic form (Akaike, 1973, Burman et al., 1994). Challenges of evaluating structured models continue to arise in applied problems (for example, Jones and Spiegelhalter, 2012).

\section{Information criteria and effective number of parameters}\label{section.waic}

For historical reasons, measures of predictive accuracy are referred to as {\em information criteria} and are typically defined based on the deviance (the log predictive density of the data given a point estimate of the fitted model, multiplied by $-2$; that is $-2\log p(y|\hat{\theta})$).

A point estimate $\hat{\theta}$ and posterior distribution $p_{\rm post}(\theta)$ are fit to the data $y$, and out-of-sample predictions will typically be less accurate than implied by the within-sample predictive accuracy.  To put it another way, the accuracy of a fitted model's predictions of future data will generally be lower, in expectation, than the accuracy of the same model's predictions for observed data---even if the family of models being fit happens to include the true data-generating process, and even if the parameters in the model happen to be sampled exactly from the specified prior distribution.

We are interested in prediction accuracy for two reasons:  first, to measure the performance of a model that we are using; second, to compare models.  Our goal in model comparison is not necessarily to pick the model with lowest estimated prediction error or even to average over candidate models---as discussed in Gelman et al.\ (2003), we prefer continuous model expansion to discrete model choice or averaging---but at least to put different models on a common scale.  Even models with completely different parameterizations can be used to predict the same measurements.

When different models have the same number of parameters estimated in the same way, one might simply compare their best-fit log predictive densities directly, but when comparing models of differing size or differing effective size (for example, comparing logistic regressions fit using uniform, spline, or Gaussian process priors), it is important to make some adjustment for the natural ability of a larger model to fit data better, even if only by chance. 

\subsection{Estimating out-of-sample predictive accuracy using available data}

Several methods are available to estimate the expected predictive accuracy without waiting for out-of-sample data.  We cannot compute formulas such as  (\ref{elppd}) directly because we do not know the true distribution, $f$.  Instead we can consider various approximations.   We know of no approximation that works in general, but predictive accuracy is important enough that it is still worth trying.  We list several reasonable-seeming approximations here.  Each of these methods has flaws, which tells us that any predictive accuracy measure that we compute will be only approximate.  
\begin{itemize}
\item {\em Within-sample predictive accuracy.}  A naive estimate of the expected log predictive density for {\em new} data is the log predictive density for {\em existing} data.  As discussed above, we would like to work with the Bayesian pointwise formula, that is, lppd as computed using the simulation (\ref{butgut3}).  This summary is quick and easy to understand but is in general an overestimate of (\ref{elppd_2}) because it is evaluated on the data from which the model was fit.
\item {\em Adjusted within-sample predictive accuracy.}  Given that lppd is a biased estimate of elppd, the next logical step is to correct that bias. Formulas such as AIC, DIC, and WAIC (all discussed below) give approximately unbiased estimates of elppd by starting with something like lppd and then subtracting a correction for the number of parameters, or the effective number of parameters, being fit.  These adjustments can give reasonable answers in many cases but have the general problem of being correct at best only in expectation, not necessarily in any given case.
\item {\em Cross-validation.}  One can attempt to capture out-of-sample prediction error by fitting the model to training data and then evaluating this predictive accuracy on a holdout set.  Cross-validation avoids the problem of overfitting but remains tied to the data at hand and thus can be correct at best only in expectation.  In addition, cross-validation can be computationally expensive:  to get a stable estimate typically requires many data partitions and fits.  At the extreme, leave-one-out cross-validation (LOO-CV) requires $n$ fits except when some computational shortcut can be used to approximate the computations.
\end{itemize}

\subsection{Akaike information criterion (AIC)}
In much of the statistical literature on predictive accuracy, inference for $\theta$  is summarized not by a posterior distribution $p_{\rm post}$ but by a point estimate $\hat{\theta}$, typically the maximum likelihood estimate. Out-of-sample predictive accuracy is then defined not by the expected log {\em posterior} predictive density (\ref{elppd}) but by $\mbox{elpd}_{\hat{\theta}}=\E_f (\log p(\tilde{y}|\hat{\theta}(y)))$ defined in (\ref{elpepd}), where both $y$ and $\tilde{y}$ are random.
There is no direct way to calculate (\ref{elpepd}); instead the standard approach is to use the log posterior density of the observed data $y$ given a point estimate $\hat{\theta}$  and correct for bias due to overfitting.

Let $k$ be the number of parameters estimated in the model.  The simplest bias correction is based on the asymptotic normal posterior distribution.  In this limit (or in the special case of a normal linear model with known variance and uniform prior distribution), subtracting $k$ from the log predictive density given the maximum likelihood estimate is a correction for how much the fitting of $k$ parameters will increase predictive accuracy, by chance alone:
\beq\label{aic}
\widehat{\mbox{elpd}}_{\rm AIC}= \log p(y|\hat{\theta}_{\rm mle}) - k.
\eeq
As defined by Akaike (1973), AIC is the above multiplied by $-2$; thus $\mbox{AIC}=-2\log p(y|\hat{\theta}_{\rm mle}) +2k$.

It makes sense to adjust the deviance for fitted parameters, but once we go beyond linear models with flat priors, we cannot simply add $k$.  Informative prior distributions and hierarchical structures tend to reduce the amount of overfitting, compared to what would happen under simple least squares or maximum likelihood estimation.

For models with informative priors or  hierarchical structure, the effective number of parameters strongly depends on the variance of the group-level parameters.  We shall illustrate in Section \ref{normalic} with the univariate normal model and in Section \ref{modelcomparison} with a classic example of educational testing experiments in 8 schools.  Under the hierarchical model in that example, we would expect the effective number of parameters to be somewhere between 8 (one for each school) and 1 (for the average of the school effects).  

There are extensions of AIC which have an adjustment related to the effective number of parameters (see Vehtari and Ojanen, 2012, section 5.5, and references therein) but these are seldom used due to stability problems and computational difficulties, issues that have motivated the construction of the more sophisticated measures discussed below.

\subsection{Deviance information criterion (DIC) and effective number of parameters}

DIC (Spiegelhalter et al., 2002) is a somewhat Bayesian version of AIC that takes formula (\ref{aic}) and makes two changes, replacing the maximum likelihood estimate $\hat{\theta}$ with the posterior mean $\hat{\theta}_{\rm Bayes}=\E(\theta|y)$ and replacing $k$ with a data-based bias correction.  The new measure of predictive accuracy is,
\beq\label{dic}
\widehat{\mbox{elpd}}_{\rm DIC}= \log p(y|\hat{\theta}_{\rm Bayes}) - p_{\rm DIC},
\eeq
where  $p_{\rm DIC}$ is the effective number of parameters, defined as,
\beq\label{dic.2}
p_{\rm DIC} = 2\left(\log p(y|\hat{\theta}_{\rm Bayes})- \E_{\rm post}(\log p(y|\theta))\right),
\eeq
where the expectation in the second term is an average of $\theta$ over its posterior distribution.  Expression (\ref{dic.2})
is calculated using simulations $\theta^s$, $s=1,\dots,S$ as,
\beq\label{pd1}
\mbox{computed } p_{\rm DIC}=  2\left(\!\log p(y|\hat{\theta}_{\rm Bayes})- \frac{1}{S}\sum_{s=1}^S \log p(y|\theta^s)\!\right).
\eeq
The posterior mean of $\theta$ will produce the maximum log predictive density when it happens to be same as the mode, and negative $p_{\rm DIC}$ can be produced if posterior mean is far from the mode.

An alternative version of DIC uses a slightly different definition of effective number of parameters:
\beq\label{pd2}
p_{\rm DIC\,alt} = 2\,\Var_{\rm post}(\log p(y|\theta)).
\eeq
Both $p_{\rm DIC}$ and $p_{\rm DIC\, alt}$ give the correct answer in the limit of fixed model and large $n$ and can be derived from the asymptotic $\chi^2$ distribution (shifted and scaled by a factor of $-\frac{1}{2}$) of the log predictive density.  For linear models with uniform prior distributions, both these measures of effective sample size reduce to $k$.  Of these two measures, $p_{\rm DIC}$ is more numerically stable but $p_{\rm DIC\, alt}$ has the advantage of always being positive. Compared to previous proposals for estimating the effective number of parameters, easier and more stable Monte Carlo approximation of DIC made it quickly popular.

The actual quantity called DIC is defined in terms of the deviance rather than the log predictive density; thus,
$$
\mbox{DIC}=-2\log p(y|\hat{\theta}_{\rm Bayes}) +2 p_{\rm DIC}.
$$

\subsection{Watanabe-Akaike information criterion (WAIC)}

WAIC (introduced by Watanabe, 2010, who calls it the widely applicable information criterion) is a more fully Bayesian approach for estimating the out-of-sample expectation (\ref{elppd_2}), starting
with the computed log pointwise posterior predictive density  (\ref{butgut3})
and then adding a correction for effective number of parameters to adjust for overfitting.

Two adjustments have been proposed in the literature.  Both are based on pointwise calculations and can be viewed as approximations to cross-validation, based on derivations not shown here.

The first approach is a difference, similar to that used to construct $p_{\rm DIC}$:
\beqnn
p_{{\rm WAIC}\, 1} = 2\sum_{i=1}^n \bigg(\!\log (\E_{\rm post} p(y_i|\theta))- \E_{\rm post}(\log p(y_i|\theta))\bigg),
\eeqnn
which can be computed from simulations by replacing the expectations by averages over the $S$ posterior draws $\theta^s$:
\beqnn
\mbox{computed } p_{{\rm WAIC}\, 1} = 2\sum_{i=1}^n\left(\log \left(\!\frac{1}{S}\sum_{s=1}^S p(y_i|\theta^s)\!\right)
 - \frac{1}{S}\sum_{s=1}^S \log p(y_i|\theta^s)\!\right)\!.
\eeqnn

The other measure uses the variance of individual terms in the log predictive density summed over the $n$ data points:
\beq\label{vlpd1.5}
p_{{\rm WAIC}\, 2} =\sum_{i=1}^n \Var_{\rm post} (\log p(y_i|\theta)).
\eeq
This expression looks similar to (\ref{pd2}), the formula for $p_{\rm DIC\, alt}$ (although without the factor of 2), but is more stable because it computes the variance separately for each data point and then sums; the summing yields stability.

To calculate (\ref{vlpd1.5}), we compute the posterior variance of the log predictive density for each data point $y_i$, that is, 
$V_{s=1}^S \log p(y_i|\theta^s)$,
where $V_{s=1}^S$ represents the sample variance, $V_{s=1}^S a_s = \frac{1}{S-1}\sum_{s=1}^S (a_s - \bar{a})^2$.
Summing over all the data points $y_i$ gives the effective number of parameters:
\beq\label{vlpd2}
\mbox{computed } p_{{\rm WAIC}\, 2} =\sum_{i=1}^n V_{s=1}^S\left( \log p(y_i|\theta^s)\right).
\eeq
We can then use either $p_{{\rm WAIC}\, 1}$ or  $p_{{\rm WAIC}\, 2}$ as a bias correction:
\beq\label{waicformula}
\widehat{\mbox{elppd}}_{\rm WAIC}= \mbox{lppd} - p_{\rm WAIC}.
\eeq

In the present article, we evaluate both $p_{\rm WAIC\,1}$ and $p_{\rm WAIC\,2}$.  For practical use, we recommend $p_{\rm WAIC\,2}$ because its series expansion has closer resemblance to the series expansion for LOO-CV and also in practice seems to give results closer to LOO-CV.

As with AIC and DIC, we define WAIC as $-2$ times the expression (\ref{waicformula}) so as to be on the deviance scale. In Watanabe's original definition, WAIC is the negative of the average log pointwise predictive density (assuming the prediction of a single new data point) and thus is divided by $n$ and does not have the factor $2$; here we scale it so as to be comparable with AIC, DIC, and other measures of deviance.


For a normal linear model with large sample size, known variance, and uniform prior distribution on the coefficients,
$p_{\rm WAIC\, 1}$ and $p_{\rm WAIC\, 2}$ are approximately equal to the number of parameters in the model.
More generally, the adjustment can be thought of as an approximation to the number of
`unconstrained' parameters in the model, where a parameter counts as
1 if it is estimated with no constraints or prior information, 0 if it
is fully constrained or if all the information about the parameter
comes from the prior distribution, or an intermediate value if both the
data and prior distributions are informative.

Compared to AIC and DIC, WAIC has the desirable property of averaging over the posterior distribution rather than conditioning on a point estimate.  This is especially relevant in a predictive context, as WAIC is evaluating the predictions that are actually being used for new data in a Bayesian context.  AIC and DIC estimate the performance of the plug-in predictive density, but Bayesian users of these measures would still use the posterior predictive density for predictions.

Other information criteria are based on Fisher's asymptotic theory assuming a regular
model for which the likelihood or the posterior converges to a single point, and where maximum likelihood and other plug-in estimates are asymptotically equivalent. WAIC works also with singular models and thus is
particularly helpful for models with hierarchical and mixture structures in which the number of parameters increases with sample size and where point estimates often do not make sense.

For all these reasons, we find WAIC more appealing than AIC and DIC.  The purpose of the present article is to gain understanding of these different approaches by applying them in some simple examples.

\subsection{Pointwise vs.\ joint predictive distribution}

A cost of using WAIC is that it relies on a partition of the data into $n$ pieces, which is not so easy to do in some structured-data settings such as time series, spatial, and network data. AIC and DIC do not make this partition explicitly, but derivations of AIC and DIC assume that residuals are independent given the point estimate $\hat{\theta}$:  conditioning on a point estimate $\hat{\theta}$ eliminates posterior dependence at the cost of not fully capturing posterior uncertainty.
Ando and Tsay (2010) have proposed an information criterion for the joint prediction, but its bias correction has the same computational difficulties as many other extensions of AIC and it can not be compared to cross-validation, since it is not possible to leave $n$ data points out in the cross-validation approach.

\subsection{Effective number of parameters as a random variable}

It makes sense that $p_{\rm DIC}$ and $p_{\rm WAIC}$ depend not just on the structure of the model but on the particular data that happen to be observed.
For a simple
example, consider the model $y_i,\dots,y_n\sim\Nor(\theta,1)$, with $n$ large and $\theta\sim
U(0,\infty)$.  That is, $\theta$ is constrained to be positive but
otherwise has a noninformative uniform prior distribution.  How many parameters
are being estimated in this model?  If the measurement $y$ is close to zero,
then the effective number of parameters $p$ is
approximately $\frac{1}{2}$, since roughly
half the information in the posterior
distribution is coming from the data and half from the prior constraint of
positivity.  However, if $y$ is positive and large, then the constraint is essentially
irrelevant, and the effective number of parameters is approximately 1.  This example illustrates that, even with a fixed model and fixed true parameters, it can make sense for the effective number of parameters to depend on data.

\subsection{`Bayesian' information criterion (BIC)}

There is also something called the Bayesian information criterion (a misleading name, we believe) that adjusts for the number of fitted parameters with a penalty that increases with the sample size, $n$ (Schwartz, 1978).  The formula is  ${\rm BIC} = -2 \log p(y|\hat{\theta}) +k \,\log n$, which for large datasets gives a larger penalty per parameter compared to AIC and thus favors simpler models.  Watanabe (2013) has also proposed a widely applicable Bayesian information criterion (WBIC) which works also in singular and unrealizable cases. BIC and its variants differ from the other information criteria considered here in being motivated not by an estimation of predictive fit but by the goal of approximating the marginal probability density of the data, $p(y)$, under the model, which can be used to estimate relative posterior probabilities in a setting of discrete model comparison.  For reasons described in Gelman and Shalizi (2012), we do not typically find it useful to think about the posterior probabilities of models but we recognize that others find BIC and similar measures helpful for both theoretical and applied reason.  For the present article, we merely point out that BIC has a different goal than the other measures we have discussed.  It is completely possible for a complicated model to predict well and have a low AIC, DIC, and WAIC, but, because of the penalty function, to have a relatively high (that is, poor) BIC.  Given that BIC is not intended to predict out-of-sample model performance but rather is designed for other purposes, we do not consider it further here.

\subsection{Leave-one-out cross-validation}
In Bayesian cross-validation, the data are repeatedly partitioned into a training set $y_{\rm train}$ and a holdout set $y_{\rm holdout}$, and then the model is fit to $y_{\rm train}$ (thus yielding a posterior distribution $p_{\rm train}(\theta)\!=\!p(\theta|y_{\rm train})$), with this fit evaluated using an estimate of the log predictive density of the holdout data, $\log p_{\rm train}(y_{\rm holdout})\!=\log \int \!p_{\rm pred}(y_{\rm holdout}|\theta)p_{\rm train}(\theta)d\theta$.  Assuming the posterior distribution $p(\theta|y_{\rm train})$ is summarized by $S$ simulation draws $\theta^s$, we calculate the log predictive density as $\log\left(\frac{1}{S}\sum_{s=1}^S p(y_{\rm holdout}|\theta^s)\right)$.

For simplicity, we will restrict our attention here to leave-one-out cross-validation (LOO-CV), the special case with $n$ partitions in which each holdout set represents a single data point.
Performing the analysis for each of the $n$ data points (or perhaps a random subset for efficient computation if $n$ is large) yields $n$ different inferences $p_{{\rm post}(-i)}$, each summarized by $S$ posterior simulations, $\theta^{is}$.

The Bayesian LOO-CV estimate of out-of-sample predictive fit is
\beqa\label{xformula1}
\label{xformula2} \mbox{lppd}_{\rm loo-cv} = \sum_{i=1}^n \log p_{{\rm post}(-i)}(y_i),\mbox{ calculated as } \sum_{i=1}^n  \log \left(\frac{1}{S}\sum_{s=1}^Sp(y_i|\theta^{is})\right).
\eeqa
Each prediction is conditioned on $n-1$ data points, which causes underestimation of the predictive fit. For large $n$ the difference is negligible, but for small $n$ (or when using $k$-fold cross-validation) we can use a first order bias correction $b$ by estimating how much better predictions would be obtained if conditioning on $n$ data points (Burman, 1989):
\beqnn
b = \mbox{lppd}-\overline{\mbox{lppd}}_{-i},
\eeqnn
where
\beqnn
\overline{\mbox{lppd}}_{-i} = \frac{1}{n} \sum_{i=1}^n\sum_{j=1}^n \log p_{{\rm post}(-i)}(y_j),\mbox{ calculated as } \frac{1}{n} \sum_{i=1}^n\sum_{j=1}^n \log\left( \frac{1}{S}\sum_{s=1}^S p(y_j|\theta^{is})\right).
\eeqnn
The bias-corrected Bayesian LOO-CV is then
\beqnn
\mbox{lppd}_{\rm cloo-cv} = \mbox{lppd}_{\rm loo-cv} + b.
\eeqnn
The bias correction $b$ is rarely used as it is usually small, but we include it for completeness.

To make comparisons to other methods, we compute an estimate of the effective
number of parameters as
\beq\label{ploocv}
p_{\rm loo-cv} = \mbox{lppd} - \mbox{lppd}_{\rm loo-cv}
\eeq
or, using bias-corrected LOO-CV,
\beqann
p_{\rm cloo-cv} &=& \mbox{lppd} - \mbox{lppd}_{\rm cloo} \\
             &=& \overline{\mbox{lppd}}_{-i} - \mbox{lppd}_{\rm loo}.
\eeqann

Cross-validation is like WAIC in that it requires data to be divided into disjoint, ideally conditionally independent, pieces.   This represents a limitation of the approach when applied to structured models.  In addition, cross-validation can be computationally expensive except in settings where shortcuts are available to approximate the distributions  $p_{{\rm post}(-i)}$ without having to re-fit the model each time. For the examples in this article such shortcuts are available, but we used the brute force approach for clarity. If no shortcuts are available, common approach is to use $k$-fold cross-validation where data is partitioned in $k$ sets. With moderate value of $k$, for example 10, computation time is reasonable in most applications.

Under some conditions, different information criteria have been shown
to be asymptotically equal to leave-one-out cross-validation (as
$n\rightarrow\infty$, the bias correction can be ignored in the proofs).  AIC has
been shown to be asymptotically equal to LOO-CV as computed using the
maximum likelihood estimate (Stone, 1997). DIC is a variation of the
regularized information criteria which have been shown to be
asymptotically equal to LOO-CV using plug-in predictive densities
(Shibata, 1989).

Bayesian cross-validation works also with singular models, and
Bayesian LOO-CV has been proven to asymptotically
equal to WAIC (Watanabe, 2010).
For finite $n$ there is a difference, as LOO-CV conditions the posterior
predictive densities on $n-1$ data points. 
These differences can be apparent for small $n$ or in hierarchical models, as we discus in our examples.

Other differences arise in regression or hierarchical models.  LOO-CV assumes the
prediction task $p(\tilde{y}_i|\tilde{x}_i,y_{-i},x_{-i})$ while WAIC
estimates $p(\tilde{y}_i|y,x)=p(\tilde{y}_i|y_i,x_i,y_{-i},x_{-i})$,
so WAIC is making predictions only at $x$-locations already observed (or in subgroups indexed by $x_i$).
This can make a noticeable difference in flexible regression models such as Gaussian processes or hierarchical models where prediction given $x_i$ may depend only weakly on all other data points $(y_{-i},x_{-i})$.  We illustrate with a simple hierarchical model in Section \ref{modelcomparison}.

The cross-validation estimates are similar to the jackknife (Efron and Tibshirani, 1993).  Even though we are working with the posterior distribution, our goal is to estimate an expectation averaging over $y^{\rm rep}$ in its true, unknown distribution, $f$; thus, we are studying the frequency properties of a Bayesian procedure.

\subsection{Comparing different estimates of out-of-sample prediction accuracy}
All the different measures discussed above are based on adjusting the log predictive density of the observed data by subtracting an approximate bias correction.  The measures differ both in their starting points and in their adjustments.

AIC starts with the log predictive density of the data conditional on the maximum likelihood estimate $\hat{\theta}$, DIC conditions on the posterior mean $\E(\theta|y)$, and WAIC starts with the log predictive density, averaging over $p_{\rm post}(\theta)=p(\theta|y)$.  Of these three approaches, only WAIC is fully Bayesian and so it is our preference when using a bias correction formula.  Cross-validation can be applied to any starting point, but it is also based on the log pointwise predictive density.

\section{Theoretical example:  normal distribution with unknown mean}\label{normalic}
In order to better understand the different information criteria, we begin by evaluating them in the context of the simplest continuous model.

\subsection{Normal data with uniform prior distribution}
Consider data $y_1,\dots,y_n\sim\Nor(\theta,1)$ with noninformative prior distribution, $p(\theta)\propto 1$.

\paragraph{AIC.}
The 
 maximum likelihood estimate is $\bar{y}$, and the probability density of the data given that estimate is 
\beq
\log p(y|\hat{\theta}_{\rm mle})=-\frac{n}{2}\log(2\pi) -\frac{1}{2}\sum_{i=1}^n(y_i-\bar{y})^2
=- \frac{n}{2}\log(2\pi) -\frac{1}{2}(n-1)s^2_y,\label{py.1}
\eeq
where $s^2_y$ is the sample variance of the data.  Only one parameter is being estimated, so
\beq
\widehat{\mbox{elpd}}_{\rm AIC} = p(y|\hat{\theta}_{\rm mle}) - k
 = -\frac{n}{2}\log(2\pi)  -\frac{1}{2}(n-1)s^2_y - 1. \label{py.1.5}
\eeq

\paragraph{DIC.}
The two pieces of DIC are $\log p(y|\hat{\theta}_{\rm Bayes})$ and the effective number of parameters $p_{\rm DIC}=2\left[\log p(y|\hat{\theta}_{\rm Bayes})- \E_{\rm post}(\log p(y|\theta))\right]$.  In this example with a flat prior density, $\hat{\theta}_{\rm Bayes}=\hat{\theta}_{\rm mle}$ and so $\log p(y|\hat{\theta}_{\rm Bayes})$ is given by (\ref{py.1}).  To compute the second term in $p_{\rm DIC}$, we start with
\beq\label{py.2}
\log p(y|\theta)=-\frac{n}{2}\log(2\pi) -\frac{1}{2}\left[n(\bar{y}-\theta)^2 + (n-1)s^2_y\right],
\eeq
and then compute the expectation of (\ref{py.2}), averaging over $\theta$ in its posterior distribution, which in this case is simply $\Nor(\theta|\bar{y},\frac{1}{n})$.  The relevant calculation is $\E_{\rm post}((\bar{y}-\theta)^2)=(\bar{y}-\bar{y})^2 + \frac{1}{n}$, and then the expectation of (\ref{py.2}) becomes,
\beq\label{py.3}
\E_{\rm post}(\log p(y|\theta))=-\frac{n}{2}\log(2\pi) -\frac{1}{2}\left[(n-1)s^2_y +1\right].
\eeq
Subtracting (\ref{py.3}) from (\ref{py.1}) and multiplying by 2 yields $p_{\rm DIC}$, which is exactly 1, as all the other terms cancel.
So, in this case, DIC and AIC are the same.

\paragraph{WAIC.}
In this example, WAIC can be easily determined analytically as well.  The first step is to write the predictive density for each data point, $p_{\rm post}(y_i)$.  In this case, $y_i|\theta\sim\Nor(\theta,1)$ and  $p_{\rm post}(\theta)=\Nor(\theta|\bar{y},\frac{1}{n})$, and so we see that $p_{\rm post}(y_i)=\Nor(y_i|\bar{y},1+\frac{1}{n})$.
Summing the terms for the $n$ data points, we get,
\beqa\nnb
\sum_{i=1}^n \log p_{\rm post}(y_i)
&=&-\,\frac{n}{2}\log(2\pi)-\frac{n}{2}\log\left(1+\frac{1}{n}\right)-\frac{1}{2}\frac{n}{n+1}\sum_{i=1}^n(y_i-\bar{y})^2\\
\label{py.4} &=&-\,\frac{n}{2}\log(2\pi)-\frac{n}{2}\log\left(1+\frac{1}{n}\right)-\frac{1}{2}\frac{n(n-1)}{n+1}s^2_y.
\eeqa

Next we determine the two forms of effective number of parameters.
To evaluate $p_{{\rm WAIC}\, 1}=2\left[\sum_{i=1}^n \log (\E_{\rm post} p(y_i|\theta))- \sum_{i=1}^n\E_{\rm post}(\log p(y_i|\theta))\right]$, the first term inside the parentheses is simply (\ref{py.4}), and the second term is
$$
\sum_{i=1}^n\E_{\rm post}(\log p(y_i|\theta)) = -\frac{n}{2}\log(2\pi)-\frac{1}{2}((n-1)s^2_y+1).
$$
Twice the difference is then,
\beq\label{py.4.5}
p_{{\rm WAIC}\, 1} = \frac{n-1}{n+1}s^2_y + 1-n\log\left(1+\frac{1}{n}\right).
\eeq

To evaluate $p_{{\rm WAIC}\, 2}=\sum_{i=1}^n \Var_{\rm post}(\log p(y_i|\theta))$, for each data point $y_i$, we start with $\log p(y_i|\theta)=\mbox{const}-\frac{1}{2}(y_i-\theta)^2$, and compute the variance of this expression, averaging over the posterior distribution, $\theta\sim\Nor(\bar{y},\frac{1}{n})$.  After the dust settles,
we get
\beq\label{py.5}
p_{{\rm WAIC}\, 2}=\frac{n-1}{n}s^2_y + \frac{1}{2n},
\eeq
and ${\rm WAIC} = -2 \sum_{i=1}^n\log p_{\rm post}(y_i) + 2 p_{\rm WAIC}$, combining (\ref{py.4}) and (\ref{py.4.5}) or (\ref{py.5}).

For this example, WAIC differs from AIC and DIC in two ways.  First, we are evaluating the pointwise predictive density averaging each term $\log p(y_i|\theta)$ over the entire posterior distribution rather than conditional on a point estimate, hence the differences between (\ref{py.1}) and (\ref{py.4}).  Second, the effective number of parameters in WAIC is not quite 1.  In the limit of large $n$, we can replace $s^2_y$ by its expected value of 1, yielding $p_{\rm WAIC}\rightarrow 1$.

WAIC is not so intuitive for small $n$.  For example, with $n=1$, the effective number of parameters $p_{\rm WAIC}$ is only 0.31 (for $p_{{\rm WAIC}\, 1}$) or 0.5 (for $p_{{\rm WAIC}\, 2}$).  As we shall see shortly, it turns out that the value 0.5 is correct, as this is all the adjustment that is needed to fix the bias in WAIC for this sample size in this example.  

\paragraph{Cross-validation.}
In this example, the leave-one-out posterior predictive densities are
\beq
p_{{\rm post}(-i)}(y_i) = \Nor\left(y_i\left|\bar{y}_{-i}, 1+\frac{1}{n-1}\right.\right),
\eeq
where $\bar{y}_{-i}$ is $\frac{1}{n-1}\sum_{j\neq i} y_j$.

The sum of the log leave-one-out posterior predictive densities is
\beq
\sum_{i=1}^n \log p_{{\rm post}(-i)}(y_i)
=-\,\frac{n}{2}\log(2\pi)-\frac{n}{2}\log\left(1+\frac{1}{n-1}\right)-\frac{1}{2}\frac{n-1}{n}\sum_{i=1}^n(y_i-\bar{y}_{-i})^2.
\eeq

For the bias correction and the effective number of parameters we need also
\beqnn
\overline{\mbox{lppd}}_{-i} = \frac{1}{n} \sum_{i=1}^n\sum_{j=1}^n \log p_{{\rm post}(-i)}(y_j) =-\,\frac{n}{2}\log(2\pi)-\frac{n}{2}\log\left(1+\frac{1}{n-1}\right)-\frac{1}{2}\frac{n-1}{n^2}\sum_{i=1}^n\sum_{j=1}^n(y_j-\bar{y}_{-i})^2.
\eeqnn

\paragraph{In expectation.}

As can be seen above, AIC, DIC, WAIC, and $\mbox{lppd}_{\rm loo-cv}$ all are random variables, in that their values depend on the data $y$, even if the model is known.  We now consider each of these in comparison to their target, the log predictive density for a new data set $\tilde{y}$.  In these evaluations we are taking expectations over both the observed data $y$ and the future data $\tilde{y}$.

Our comparison point is the expected log pointwise predictive density (\ref{elppd_2}) for new data:
\beqnn
\mbox{elppd} =  \sum_{i=1}^n \E(\log p_{\rm post}(\tilde{y}_i))
= -\,\frac{n}{2}\log(2\pi)-\frac{n}{2}\log\left(1+\frac{1}{n}\right)-\frac{1}{2}\frac{n}{n+1}\sum_{i=1}^n\E\left((\tilde{y}_i-\bar{y})^2\right).
\eeqnn
This last term can be decomposed and evaluated:
\beqa\nnb
\E\left(\sum_{i=1}^n(\tilde{y}_i-\bar{y})^2\right) &=& (n-1)\E (s^2_{\tilde y}) + n\E((\overline{\tilde{y}}-\bar{y})^2)\\
&=& n+1,
\label{tildesum}
\eeqa
and thus the the expected log pointwise predictive density is
\beq
\mbox{elppd} 
= -\,\frac{n}{2}\log(2\pi)-\frac{n}{2}\log\left(1+\frac{1}{n}\right)-\frac{n}{2}.
\label{py.6}
\eeq

We also need the expected value of the log pointwise predictive density for existing data, which can be obtained by plugging $\E(s^2_y)=1$ into (\ref{py.4}):
\beq\label{py.6.5}
\E(\mbox{lppd})= -\,\frac{n}{2}\log(2\pi)-\frac{n}{2}\log\left(1+\frac{1}{n}\right)-\frac{1}{2}\frac{n(n-1)}{n+1}.
\eeq
Despite what the notation might seem to imply, elppd is {\em not} the same as $\E(\mbox{lppd})$; the former is the expected log pointwise predictive density for future data $\tilde{y}$, while the latter is this density evaluated at the observed data $y$.

The correct `effective number of parameters' (or bias correction) is the difference between $\E(\mbox{lppd})$ and $\mbox{elppd}$, that is, from (\ref{py.6}) and (\ref{py.6.5}),
\beq
\label{true.p}
\E(\mbox{lppd}) - \mbox{elppd} = \frac{n}{2} - \frac{1}{2} \frac{n(n-1)}{n+1} = \frac{n}{n+1},
\eeq
which is always less than 1:  with $n=1$ it is 0.5 and in the limit of large $n$ it goes to 1.

The target of AIC and DIC is the performance of the plug-in predictive density. Thus for comparison we also calculate
\beqnn
\E\left[\log p(\tilde{y}|\hat{\theta}(y))\right] =\E\left[\log \prod_{i=1}^n\Nor(\tilde{y}_i|\bar{y},1)\right]
=-\frac{n}{2}\log(2\pi) -  \frac{1}{2}\,\E\left[\sum_{i=1}^n(\tilde{y}_i-\bar{y})^2\right].
\eeqnn
Inserting (\ref{tildesum}) into that last term yields the expected out-of-sample log density given the point estimate as
\beqnn
 \E\left(\log p(\tilde{y}|\hat{\theta}(y))\right) = \frac{n}{2}\log(2\pi) - \frac{n}{2} -\frac{1}{2}.
\eeqnn

\paragraph{In expectation:  AIC and DIC.}

The expectation of AIC from (\ref{py.1.5}) is,
\beqnn
\E(\widehat{\mbox{elpd}}_{\rm AIC}) = -\frac{n}{2}\log(2\pi)  - \frac{1}{2}(n-1)\E(s^2_y) - 1
 = -\frac{n}{2}\log(2\pi) - \frac{n}{2} - \frac{1}{2},
\eeqnn
and also for DIC, which in this simple noninformative normal example is the same as AIC.  Thus AIC and DIC unbiasedly estimate the log predictive density given point estimate for new data for this example.

We can subtract the above expected estimate from its target, expression (\ref{py.6}), to obtain:
\beqnn
\mbox{elppd}-\E(\widehat{\mbox{elpd}}_{\rm AIC})
 = -\frac{n}{2}\log\left(1+\frac{1}{n}\right) + \frac{1}{2}
 = \frac{1}{2n}\frac{n-\frac{2}{3}}{2n}+o(n^{-3})
 = \frac{1}{4n}+o(n^{-2}),
\eeqnn
In this simple example, the estimated effective number of parameters differs from the appropriate expectation (\ref{true.p}), but combining two wrongs makes a right, and AIC/DIC performs decently.

\paragraph{In expectation:  WAIC.}
We can obtain the expected values of the two versions of $p_{\rm WAIC}$, by taking expectations of (\ref{py.4.5}) and (\ref{py.5}), to yield,
\beqnn
\E(p_{{\rm WAIC}\, 1})=\frac{n-1}{n+1}+1-n\log(1+\frac{1}{n})=\frac{n-\frac{1}{2}+\frac{1}{6n}}{n+1} + o(n^{-3})
\eeqnn
 and
\beqnn
\E(p_{{\rm WAIC}\, 2})=1-\frac{1}{2n}=\frac{n-\frac{1}{2}}{n}.
\eeqnn
For large $n$, the limits work out; the difference (\ref{true.p}) and both versions of $p_{\rm WAIC}$ all approach 1, which is appropriate for this example of a single parameter with noninformative prior distribution.  At the other extreme of $n\!=\!1$, the difference (\ref{true.p}) and $\E(p_{{\rm WAIC}\, 2})$ take on the value $\frac{1}{2}$, while $\E(p_{{\rm WAIC}\, 1})$ is slightly off with a value of 0.31. Asymptotic errors are
\beqann
\mbox{elppd}-\E(\widehat{\mbox{elppd}}_{{\rm WAIC}\, 1}) &=& \frac{1}{2n}\frac{n-\frac{1}{3}}{n+1} + o(n^{-3})= \frac{1}{2n+2} +o(n^{-2}) \\
\mbox{elppd}-\E(\widehat{\mbox{elppd}}_{{\rm WAIC}\, 2}) &=& -\frac{1}{2n}\frac{n-1}{n+1} = -\frac{1}{2n+2} +o(n^{-2}).
\eeqann

\paragraph{In expectation: Cross-validation.}

The expectation over $y$ is,
\beqnn
\widehat{\mbox{elppd}}_{\rm loo-cv}
=-\,\frac{n}{2}\log(2\pi)-\frac{n}{2}\log\left(1+\frac{1}{n-1}\right)-\frac{1}{2}\frac{n-1}{n}\sum_{i=1}^n\E\left((y_i-\bar{y}_{-i})^2\right). 
\eeqnn
Terms $(y_i-\bar{y}_{-i})^2$ are not completely independent as $\bar{y}_{-i}$ are overlapping, but it does not affect the expectation.
Using (\ref{tildesum}) to evaluate the sum in the last term we get
\beqnn
\sum_{i=1}^n\E\left((y_i-\bar{y}_{-i})^2\right)=n+\frac{n}{n-1}
\eeqnn
and
\beqnn
\E(\widehat{\mbox{elppd}}_{\rm loo-cv}) = -\,\frac{n}{2}\log(2\pi)-\frac{n}{2}\log\left(1+\frac{1}{n-1}\right)-\frac{n}{2}.
\eeqnn
Subtracting this from elppd yields, for $n>1$,
\beqnn
\mbox{elppd}-\E(\widehat{\mbox{elppd}}_{\rm loo-cv}) = -\frac{n}{2}\log\left(1+\frac{1}{n}\right)+\frac{n}{2}\log\left(1+\frac{1}{n-1}\right)
= -\frac{n}{2}\log\left(1-\frac{1}{n^2}\right) 
= \frac{1}{2n}+o(n^{-3}).
\eeqnn
This difference comes from the fact the cross-validation conditions on $n-1$ data points.

For the bias correction we need
\beqnn
\E(\overline{\mbox{lppd}}_{-i})=-\,\frac{n}{2}\log(2\pi)-\frac{n}{2}\log\left(1+\frac{1}{n-1}\right) - \frac{n}{2} + 1 - \frac{1}{n}.
\eeqnn
The bias-corrected LOO-CV is
\beqnn
\E(\widehat{\mbox{elppd}}_{\rm cloo-cv})  = -\,\frac{n}{2}\log(2\pi)-\frac{n}{2}\log\left(1+\frac{1}{n}\right)-\frac{n}{2}+\frac{1}{n^2+n}.
\eeqnn
Subtracting this from elppd yields,
\beqnn
\mbox{elppd}-\E(\widehat{\mbox{elppd}}_{\rm cloo-cv}) = -\,\frac{1}{n^2+n},
\eeqnn
showing much improved accuracy from the bias correction.


The effective number of parameters from leave-one-out cross-validation is, for $n>1$,
\beqann
p_{\rm loo-cv}&=&\E(\mbox{lppd})-\E(\mbox{lppd}_{\rm loo-cv})\\
&=& -\frac{n}{2}\log\left(1+\frac{1}{n}\right)+\frac{n}{2}\log\left(1+\frac{1}{n-1}\right) -\frac{1}{2}\frac{n(n-1)}{n+1} + \frac{n}{2}\\
&=& -\frac{n}{2}\log\left(1-\frac{1}{n^2}\right) + \frac{n}{n+1} \\
&=& \frac{n+\frac{1}{2}+\frac{1}{2n}}{n+1} +o(n^{-3})
\eeqann
and, from the-bias corrected version:
\beqnn
p_{\rm cloo-cv}=\E(\mbox{lppd})-\E(\mbox{lppd}_{\rm cloo-cv})
= \frac{n-1}{n}.
\eeqnn


\subsection{Normal data with informative prior distribution}
The above calculations get more interesting when we add prior information so that, in effect, less than a full parameter is estimated from any finite data set.  We consider data $y_1,\dots,y_n\sim\Nor(\theta,1)$  with a normal prior distribution, $\theta\sim\Nor(\mu,\tau^2)$. To simplify the algebraic expressions we shall write $m=\frac{1}{\tau^2}$, the prior precision and equivalent number of data points in the prior.  The posterior distribution is then $p_{\rm post}(\theta)=\Nor(\theta|\frac{m\mu+n\bar{y}}{m+n},\frac{1}{m+n})$ and the posterior predictive distribution for a data point $y_i$ is $p_{\rm post}(y_i) =\Nor(y_i|\frac{m\mu+n\bar{y}}{m+n},1+\frac{1}{m+n})$.

\paragraph{AIC.}  Adding a prior distribution does not affect the maximum likelihood estimate, so $\log p(y|\hat{\theta}_{\rm mle})$ is unchanged from (\ref{py.1}), and AIC is the same as before.

\paragraph{DIC.} The posterior mean is $\hat{\theta}_{\rm Bayes}=\frac{m\mu+n\bar{y}}{m+n}$, and so and the first term in DIC is 
\beqa\nnb
\!\!\!\!\!\!\!\!\!\!\!\!\!\!\!\!\!\!\!\log p(y|\hat{\theta}_{\rm Bayes})&\!\!=\!\!&- \,\frac{n}{2}\log(2\pi) -\frac{1}{2}(n-1)s^2_y-\frac{1}{2}n(\bar{y}-\hat{\theta}_{\rm Bayes})^2\\
&\!\!=\!\!&- \,\frac{n}{2}\log(2\pi) -\frac{1}{2}(n-1)s^2_y-\frac{1}{2}n\left(\frac{m}{m+n}\right)^2\!\!(\bar{y}-\mu)^2.\label{py.7}
\eeqa
Next we evaluate (\ref{dic.2}) for this example; after working through the algebra, we get,
\beq\label{py.8}
p_{\rm DIC}=\frac{n}{m+n},
\eeq
which makes sense:  the flat prior corresponds to $m\!=\!0$, so that $p_{\rm DIC}\!=\!1$; at the other extreme, large values of $m$ correspond to prior distributions that are much more informative than the data, and $p_{\rm DIC}\rightarrow 0$.

\paragraph{WAIC.}
Going through the algebra, the log pointwise predictive density of the data is
\beqann
{\rm lppd}&=& \sum_{i=1}^n \log p_{\rm post}(y_i)\\
& =& -\,\frac{n}{2}\log(2\pi) - \frac{n}{2}\log\left(1+\frac{1}{m+n}\right)-\frac{1}{2}\frac{(m+n)(n-1)}{(m+n+1)}s^2_y  - \frac{1}{2}\frac{m^2}{(m+n)(m+n+1)} (\bar{y}-\mu)^2.
\eeqann
We can also work out
\beqann
p_{{\rm WAIC}\, 1} &=&
\frac{n-1}{m+n+1}s^2_y  + \frac{m^2}{(m+n)^2(m+n+1)} (\bar{y}-\mu)^2 + \frac{n}{m+n} -  n\log\left(1+\frac{1}{m+n}\right) \\
p_{{\rm WAIC}\, 2} &=& \frac{n-1}{m+n}s^2_y + \frac{m^2n}{(m+n)^3}(\bar{y}-\mu)^2+\frac{n}{2(m+n)^2}.
\eeqann
We can understand these formulas by applying them to special cases:
\begin{itemize}
\item A flat prior distribution ($m\!=\!0$) yields  $p_{{\rm WAIC}\, 1}=  \frac{n-1}{n+1}s^2_y +1-n\log\left(1+\frac{1}{n}\right)$ and $p_{{\rm WAIC}\, 2}=\frac{n-1}{n}s^2_y + \frac{1}{2n}$, same as (\ref{py.4.5}) and  (\ref{py.5}).  In expectation, these are
 $\E(p_{{\rm WAIC}\, 1})=\frac{n-1}{n+1}+1-n\log(1+\frac{1}{n})$ and  $\E(p_{{\rm WAIC}\, 2})=1-\frac{1}{2n}$, as before. 
\item A prior distribution equally informative as data ($m\!=\!n$) yields $p_{{\rm WAIC}\, 1}=\frac{n-1}{2n+1 }s^2_y+\frac{1}{4(2n+1)}(\bar{y}-\mu)^2 + \frac{1}{2}-n\log(1+\frac{1}{2n})$ and
$p_{{\rm WAIC}\, 2}=\frac{n-1}{2n}s^2_y + \frac{1}{8}(\bar{y}-\mu)^2 + \frac{1}{8n}$.  In expectation, averaging over the prior distribution and the data model, these both look like $\frac{1}{2}-o(n)$, approximately halving the effective number of parameters, on average, compared to the noninformative prior.
\item A completely informative prior distribution ($m\!=\!\infty$) yields $p_{\rm WAIC}=0$, which makes sense, as the data provide no information.
\end{itemize}

\paragraph{Cross-validation.}
For LOO-CV, we need
\beqnn
p_{{\rm post}(-i)}(y_i)=\Nor\left(y_i\left|\frac{m\mu+(n-1)\bar{y}_{-i}}{m+n-1},1+\frac{1}{m+n-1}\right.\right)
\eeqnn
and
\beqann
\sum_{i=1}^n \log p_{{\rm post}(-i)}(y_i)&=& -\frac{n}{2}\log(2\pi)-\frac{n}{2}\log\left(1+\frac{1}{m+n-1}\right)\\
&&-\,\frac{1}{2}\frac{m+n-1}{m+n}\sum_{i=1}^n\left(y_i-\frac{m\mu+(n-1)\bar{y}_{-i}}{m+n-1}\right)^2.
\eeqann
The expectation of the sum in the the last term is (using the marginal expectation, $\E(y)=\mu$), 
\beqnn
\E\left(y_i-\frac{m\mu+(n-1)\bar{y}_{-i}}{m+n-1}\right)^2 = n + \frac{n(n-1)}{(m+n-1)^2},
\eeqnn
and the expectation of leave-one-out cross-validation is
\beqnn
\E(\mbox{lppd}_{\rm cloo-cv}) = -\frac{n}{2}\log(2\pi)-\frac{n}{2}\log\left(1+\frac{1}{m+n-1}\right)-\frac{n}{2}\left(\frac{(m+n-1)^2+n-1}{(m+n)(m+n-1)}\right).
\eeqnn
If $m=0$ this is same as with uniform prior and it increases with increasing $m$.

\subsection{Hierarchical normal model}
Next consider the balanced model with data $y_{ij}\sim\Nor(\theta_j,1), \mbox{ for } i=1,\dots,n;\, j=1,\dots,J$, and prior distribution $\theta_j\sim\Nor(\mu,\tau^2)$.  If the hyperparameters are known, the inference reduces to the non-hierarchical setting described above:  the $J$ parameters have independent normal posterior distributions, each based on $J$ data points.  AIC and DIC are unchanged except that the log predictive probabilities and effective numbers of parameters are summed over the groups.  The results become more stable (because we are averaging $s^2_y$ for $J$ groups) but the algebra is no different.

For WAIC and cross-validation, though, there is a possible change, depending on how the data are counted.  If each of the $nJ$ observations is counted as a separate data point, the results again reduce to $J$ copies of what we had before.  But another option is to count each {\em group} as a separate data point, in which case the log pointwise predictive density (\ref{butgut3}) changes, as does $p_{\rm WAIC}$ in (\ref{vlpd1.5}) or (\ref{vlpd2}), as all of these now are average over $J$ larger terms corresponding to the $J$ vectors $y_j$, rather than $nJ$ little terms corresponding to the individual $y_{ij}$'s.

In a full hierarchical model with hyperparameters unknown, DIC and WAIC both change in recognition of this new source of posterior uncertainty (while AIC is not clearly defined in such cases).  We illustrate in section \ref{modelcomparison}, evaluating these information criteria for a hierarchical model with {\em unknown} hyperparameters, as fit to the data from the 8-schools study.

\section{Simple applied example:  election forecasting}

\begin{figure}
\centerline{\psfig{figure=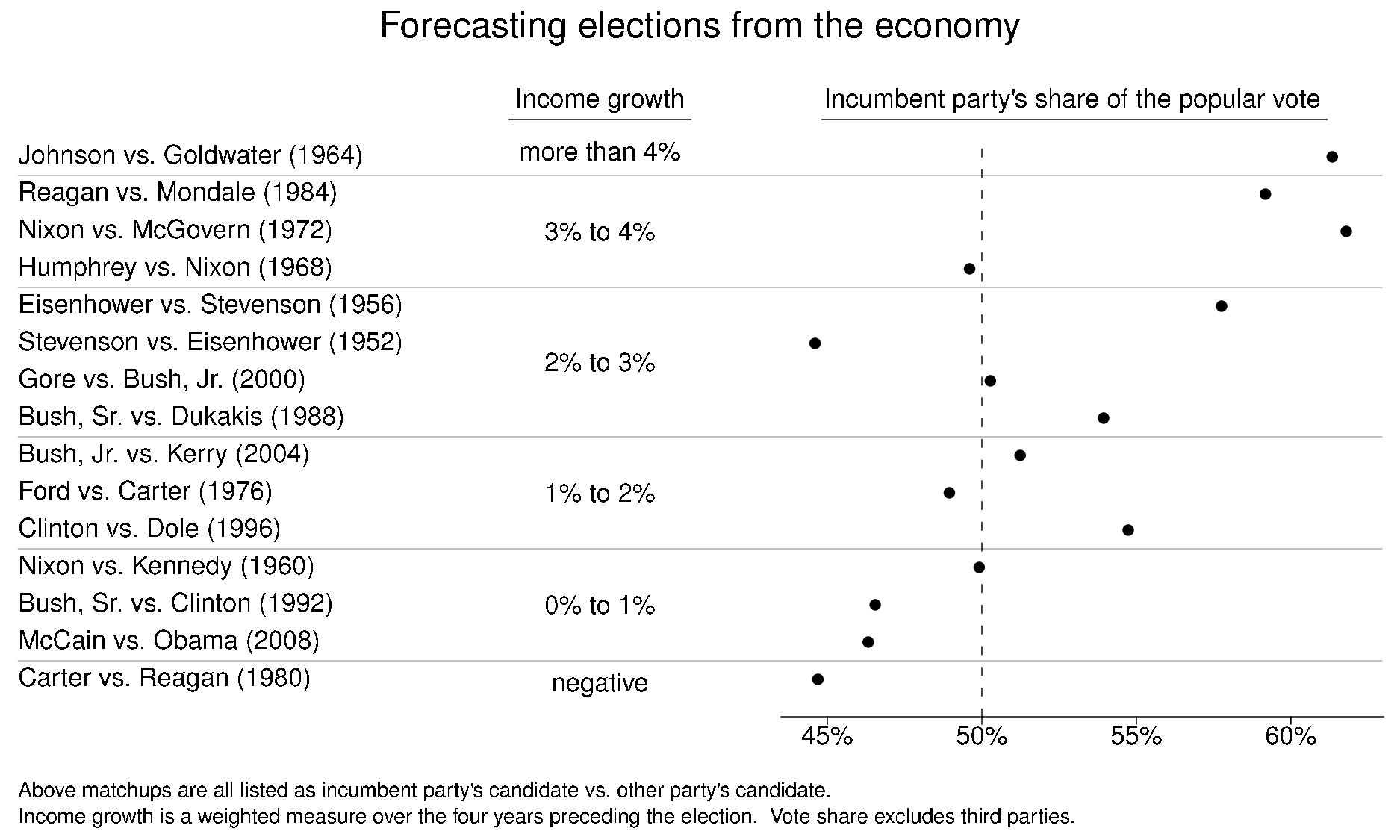,width=.85\textwidth}}
\caption{\em Douglas Hibbs's `bread and peace' model of voting and the economy. Presidential elections since 1952 are listed in order of the economic performance at the end of the preceding administration (as measured by inflation-adjusted growth in average personal income). The better the economy, the better the incumbent party's candidate generally does, with the biggest exceptions being 1952 (Korean War) and 1968 (Vietnam War).}
\label{hibbs1}
\end{figure}

We illustrate the ideas using a simple linear prediction problem.   Figure \ref{hibbs1} shows a quick summary of economic conditions and presidential elections over the past several decades. It is based on the `bread and peace' model created by political scientist Douglas Hibbs (see Hibbs, 2008, for a recent review) to forecast elections based solely on economic growth (with corrections for wartime, notably
Adlai Stevenson's exceptionally poor performance in 1952 and Hubert Humphrey's loss in 1968, years when Democrats were presiding over unpopular wars). Better forecasts are possible using additional information such as incumbency and opinion polls, but what is impressive here is that this simple model does pretty well all by itself.

For simplicity, we predict $y$ (vote share) solely from $x$ (economic performance), using a linear regression, $y\sim\Nor(a+b x,\sigma^2)$, with a noninformative prior distribution, $p(a,b,\log\sigma)\propto 1$, so that the posterior distribution is normal-inverse-$\chi^2$.
Fit to all 15 data points in Figure \ref{hibbs1}, the posterior mode $(\hat{a},\hat{b},\hat{\sigma})$ is $(45.9,3.2,3.6)$.  Although these data form a time series, we are treating them here as a simple regression problem.  In particular, when considering leave-one-out cross-validation, we do not limit ourselves to predicting from the past; rather, we consider the elections as 15 independent data points.

\paragraph{Posterior distribution of the observed log predictive density, $p(y|\theta)$.}  

\begin{figure}
\centerline{\psfig{figure=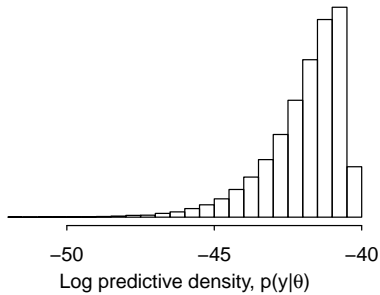,height=2.5in,angle=270}}
\caption{\em Posterior distribution of the log predictive density $\log p(y|\theta)$ for the election forecasting example.  The variation comes from posterior uncertainty in $\theta$.  The maximum value of the distribution, $-40.3$, is the log predictive density when $\theta$ is at the maximum likelihood estimate.  The mean of the distribution is $-42.0$, and the difference between the mean and the maximum is 1.7, which is close to the value of $\frac{3}{2}$ that would be predicted from asymptotic theory, given that we are estimating 3 parameters (two coefficients and a residual variance).}
\label{llsim}
\end{figure}

In our regression example, the log predictive probability density of the data is $\sum_{i=1}^{15}\log(\Nor(y_i|a+bx_i,\sigma^2))$, with an uncertainty induced by the posterior distribution, $p_{\rm post}(a,b,\sigma^2)$.  The posterior distribution $p_{\rm post}(\theta)=p(a,b,\sigma^2|y)$ is normal-inverse-$\chi^2$.  To get a sense of uncertainty in the log predictive density $p(y_i|\theta)$, we compute it for each of $S=10,\!000$ posterior simulation draws of $\theta$.  Figure \ref{llsim} shows the resulting distribution, which looks roughly like a $\chi^2_3$ (no surprise since three parameters are being estimated---two coefficients and a variance---and the sample size of 15 is large enough that we would expect the asymptotic normal approximation to the posterior distribution to be pretty good), scaled by a factor of $-\frac{1}{2}$ and shifted so that its upper limit corresponds to the maximum likelihood estimate (with log predictive density of $-40.3$, as noted earlier).  The mean of the posterior distribution of the log predictive density is $-42.0$, and the difference between the mean and the maximum is 1.7, which is close to the value of $\frac{3}{2}$ that would be predicted from asymptotic theory, given that 3 parameters are being estimated.

Figure \ref{llsim} is reminiscent of the direct likelihood methods of Dempster (1974) and Aitkin (2010).  Our approach is different, however, in being fully Bayesian, with the apparent correspondence appearing here only because we happen to be using a flat prior distribution.  A change in $p(\theta)$ would result in a different posterior distribution for $p(y|\theta)$ and thus a different Figure \ref{llsim}, a different expected value, and so forth.

\paragraph{Log predictive density of the observed data.}   For this simple linear model, the posterior predictive distribution of any data point has an analytic form ($t$ with $n\!-\!1$ degrees of freedom), but it is easy enough to use the more general simulation-based computational formula (\ref{butgut3}).  Calculated either way, the log pointwise predictive density is ${\rm lppd}=-40.9$.  Unsurprisingly, this number is slightly lower than the predictive density evaluated at the maximum likelihood estimate:  averaging over uncertainty in the parameters yields a slightly lower probability for the observed data.

\paragraph{AIC.}
Fit to all 15 data points, the MLE $(\hat{a}, \hat{b}, \hat{\sigma})$ is $(45.9,3.2,3.6)$. Since 3 parameters are estimated, the value of $\widehat{\textnormal{elpd}}_{\mathrm{AIC}}$ is
\[\sum_{i=1}^{15} \log \textnormal{N}(y_i | 45.9 + 3.2 x_i, 3.6^2) \,- 3 = -\,43.3,\]
and $\mbox{AIC} =-2 \,\widehat{\textnormal{elpd}}_{\mathrm{AIC}} = 86.6.$

\paragraph{DIC.}
The relevant formula is
$p_{\mathrm{DIC}} = 2 \left(\log p(y|\E_\mathrm{post}(\theta)) - \E_\mathrm{post}(\log p(y|\theta)) \right)$.

The second of these terms is invariant to reparameterization; we calculate it as 
\[\E_\mathrm{post}(y|\theta) = \frac{1}{S}\sum_{s=1}^S \sum_{i=1}^{15} \log \textnormal{N}(y_i|a^s + b^s x_i, (\sigma^s)^2)=-42.0,\]
based on a large number  $S$ of simulation draws.

The first term is not invariant. With respect to the prior $p(a,b,\log\sigma) \propto 1$, the posterior means of $a$ and $b$ are $45.9$ and $3.2$, the same as the maximum likelihood estimate. The posterior means of $\sigma$, $\sigma^2$, and $\log\sigma$ are $\E(\sigma|y) = 4.1$, $\E(\sigma^2|y) = 17.2$, and $\E(\log\sigma | y) = 1.4$.
Parameterizing using $\sigma$, we get
\[\log p(y|\E_\mathrm{post}(\theta)) = \sum_{i=1}^{15} \log \textnormal{N}(y_i | \E(a|y) + \E(b|y) x_i, (\E(\sigma|y))^2)=-40.5,\]
which gives $p_\mathrm{DIC} = 2(-40.5 - (-42.0)) = 3.0$, $\widehat{\textnormal{elpd}}_\mathrm{DIC} = \log p(y|\E_\mathrm{post}(\theta)) - p_\mathrm{DIC} = -40.5 - 3.0 = -43.5$, and $\textnormal{DIC} = -2\,\widehat{\textnormal{elpd}}_\mathrm{DIC} = 87.0$.

\paragraph{WAIC.} 
The log pointwise predictive probability of the observed data under the fitted model is
$$
\mbox{lppd} =  \sum_{i=1}^{15}\log \left(\frac{1}{S} \sum_{s=1}^S \textnormal{N}(y_i | a^s + b^s x_i, (\sigma^s)^2) \right) = -40.9.
$$
The effective number of parameters can be calculated as
\[p_{{\rm WAIC}\, 1} = 2\big(\mbox{lppd} - \E_{\rm post}(y|\theta)\big),
= 2(-40.9-(-42.0)) = 2.2\]
or
\[p_{{\rm WAIC}\, 2} = \sum_{i=1}^{15} V_{s=1}^S \log \textnormal{N}(y_i | a^s + b^s x_i, (\sigma^s)^2)=2.7.\]
Then
$\widehat{\textnormal{elppd}}_{{\rm WAIC}\, 1} = \mbox{lppd} - p_{{\rm WAIC}\, 1} = -40.9-2.2=-43.1$,
and $\widehat{\textnormal{elppd}}_{{\rm WAIC}\,  2} = \mbox{lppd} - p_{{\rm WAIC}\, 2} = -40.9-2.7=-43.6$,
so $\mbox{WAIC}$ is 86.2 or 87.2.

\paragraph{Leave-one-out cross-validation.} 
We fit the model 15 times, leaving out a different data point each time. For each fit of the model, we sample $S$ times from the posterior distribution of the parameters and compute the log predictive density. The cross-validated pointwise predictive accuracy is
\[{\rm lppd}_\mathrm{loo-cv} = \sum_{l=1}^{15} \log \left( \frac{1}{S} \sum_{s=1}^S \textnormal{N}(y_l | a^{ls} + b^{ls} x_l, (\sigma^{ls})^2) \right),\]
which equals $-43.8$.  Multiplying by $-2$ to be on the same scale as AIC and the others, we get 87.6.
The effective number of parameters from cross-validation, from (\ref{ploocv}), is $p_{\rm loo-cv} = \E(\mbox{lppd}) - \E(\mbox{lppd}_{\rm loo-cv})-40.9 - (-43.8)=2.9$.

Given that this model includes two linear coefficients and a variance parameter, these all look reasonable as an effective number of parameters. 

\section{Simple applied example:  meta-analysis of educational testing experiments}\label{modelcomparison}

\begin{table}
\small{
\begin{center}\btab{ccc}
&\multicolumn{1}{c}{Estimated}&\multicolumn{1}{c}{Standard error}\\
&\multicolumn{1}{c}{treatment}&\multicolumn{1}{c}{of effect}\\
School&\multicolumn{1}{c}{effect, $y_j$}&
\multicolumn{1}{c}{estimate, $\sigma_j$}\\\hline
A& \ 28 & 15 \\
B& \ \,\, 8 & 10 \\
C& $\,-3$ & 16 \\
D& \ \,\, 7 & 11 \\
E& $\,-1$ & \ 9 \\
F& \ \,\, 1 & 11 \\
G& \ 18 & 10 \\
H& \ 12 & 18
\etab\end{center}
}
\caption{Observed effects of special
preparation on test scores in eight randomized
experiments.  Estimates are based on separate analyses for the
eight experiments. From Rubin (1981).}\label{tab5.2}
\end{table}

We next explore Bayesian predictive error models in the context of a classic example from Rubin (1981) of an educational testing experiment, measuring the effects of a test preparation program performed in eight different high schools in New Jersey. A separate randomized experiment was conducted in each school, and the administrators of each school implemented the program in their own way.  The results, based on a separate regression analyses performed in each school, are displayed in Table \ref{tab5.2}.  Three modes of inference were proposed for these data:
\begin{itemize}
\item {\em No pooling}: Separate estimates for each of the eight schools, reflecting that the experiments were performed independently and so each school's observed value is an unbiased estimate of its own treatment effect.  This model has eight parameters:  an estimate for each school.
\item {\em Complete pooling}:  A combined estimate averaging the data from all schools into a single number, reflecting that the eight schools were actually quite similar (as were the eight different treatments), and also reflecting that the variation among the eight estimates (the left column of numbers in Table \ref{tab5.2}) is no larger than would be expected by chance alone given the standard errors (the rightmost column in the table). This model has only one, shared, parameter.
\item {\em Hierarchical model:}  A Bayesian meta-analysis, partially pooling the eight estimates toward a common mean.  This model has eight parameters but they are constrained through their hierarchical distribution and are not estimated independently; thus the effective number of parameters should be some number less than 8.
\end{itemize}
Rubin (1981) used this small example to demonstrate the feasibility and benefits of a full Bayesian analysis, averaging over all parameters and hyperparameters in the model.  Here we shall take the Bayesian model as given.  We throw this example at the predictive error measures because it is a much-studied and well-understood example, hence a good way to develop some intuition about the behavior of AIC, DIC, WAIC and cross-validation in a hierarchical setting.

\begin{figure}
\vspace{-.15in}
\centerline{\psfig{figure=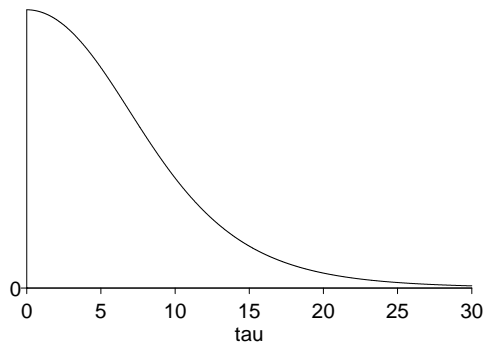,height=2.7in,angle=270}}
\vspace{-.05in}
\caption{Marginal posterior density, $p(\tau|y)$, for standard deviation
of the population of school effects $\theta_j$ in the
educational testing example.}
\label{fig5.5}
\end{figure}

The hierarchical model is $y_j\sim\Nor(\theta_j,\sigma_j^2),\, \theta_j\sim\Nor(\mu,\tau^2), \mbox{ for } j=1,\dots,J$, where $y_j$ and $\sigma_j$ are the estimate and standard error for the treatment effect in school $j$, and the hyperparameters $\mu,\tau$ determine the population distribution of the effects in the schools.  We assume a uniform hyperprior density, $p(\mu,\tau)\propto 1$, and the resulting posterior distribution for the group-level scale parameter $\tau$ is displayed in Figure \ref{fig5.5}.  The posterior mass is concentrated near 0, implying that the evidence is that there is little variation in the true treatment effects across the $J=8$ schools.

Table\label{8schools.deviance}
\ref{deviance.table} illustrates the use of predictive log densities and information criteria to compare
the three models---no pooling, complete pooling, and
hierarchical---fitted to the SAT coaching data.  We only have data at the group level, so we necessarily define our data points and cross-validation based on the 8 schools, not the individual students.

\begin{table}
\small{
\begin{center}
\btab{ll rrr}
 && \multicolumn{1}{c}{No} &  \multicolumn{1}{c}{Complete} &  \multicolumn{1}{c}{Hierarchical} \\
 &&  \multicolumn{1}{c}{pooling} &  \multicolumn{1}{c}{pooling} &  \multicolumn{1}{c}{model} \\
 &&  \multicolumn{1}{c}{($\tau=\infty$)} &  \multicolumn{1}{c}{($\tau=0$)} &  \multicolumn{1}{c}{($\tau$ estimated)} \\\hline
 & $-2\,\mbox{lpd} = -2 \log p(y|\hat{\theta}_{\rm mle})$ &$54.6$ \ \ \ & $59.4$ \ \ \ & \ \ \ \ \ \ \\
AIC & $k$  & 8.0 \ \ \ & 1.0 \ \ \ & \ \ \ \ \ \ \\
 & ${\rm AIC}=-2\,\widehat{\rm elpd}_{\rm AIC}$  & $70.6$ \ \ \ & $61.4$ \ \ \ & \ \ \ \ \ \ \\\hline
 & $-2\,\mbox{lpd} = -2\log p(y|\hat{\theta}_{\rm Bayes})$  & $54.6$ \ \ \ & $59.4$ \ \ \ & $\!\!\!\!\! 57.4$ \ \ \ \ \ \ \\
DIC & $p_{\rm DIC}$  &8.0 \ \ \ &1.0 \ \ \ & $\!\!\!\!\! 2.8$ \ \ \ \ \ \ \\
 & ${\rm DIC}=-2\,\widehat{\rm elpd}_{\rm DIC}$  &$70.6$ \ \ \ & $61.4$ \ \ \ & $\!\!\!\!\! 63.0$ \ \ \ \ \ \ \\\hline
 & $-2\,\mbox{lppd} = -2\sum_i\log p_{\rm post}(y_i)$  & $60.2$ \ \ \ & $59.8$ \ \ \ & $\!\!\!\!\! 59.2$ \ \ \ \ \ \ \\
WAIC & $p_{{\rm WAIC}\, 1}$  & 2.5 \ \ \ & 0.6 \ \ \ & $\!\!\!\!\! 1.0$ \ \ \ \ \ \ \\
           & $p_{{\rm WAIC}\, 2}$  & 4.0 \ \ \ & 0.7 \ \ \ & $\!\!\!\!\! 1.3$ \ \ \ \ \ \ \\
 & ${\rm WAIC}=-2\,\widehat{\rm elppd}_{{\rm WAIC}\, 2}$  & $68.2$ \ \ \ & $61.2$ \ \ \ & $\!\!\!\!\! 61.8$ \ \ \ \ \ \ \\\hline
 & $-2\,\mbox{lppd}$  &  \ \ \ & $59.8$ \ \ \ & $\!\!\!\!\! 59.2$ \ \ \ \ \ \ \\
LOO-CV & $p_{\rm loo-cv}$  & \ \ \ &  0.5 \ \ \ & $\!\!\!\!\! 1.8 $ \ \ \ \ \ \ \\
& $-2\,{\rm lppd}_{\rm loo-cv}$  & \ \ \ & $60.8$ \ \ \ & $\!\!\!\!\! 62.8$ \ \ \ \ \ \ 
\etab
\end{center}
}
\caption{\em Deviance (-2 times log predictive density) and corrections for parameter fitting using AIC, DIC, WAIC (using the correction $p_{{\rm WAIC}\, 2}$), and leave-one-out cross-validation
for each of three models fitted to the data in Table \ref{tab5.2}.  Lower values of AIC/DIC/WAIC imply higher predictive accuracy.
\newline
Blank cells in the table correspond to measures that are undefined:  AIC is defined relative to the maximum likelihood estimate and so is inappropriate for the hierarchical model; cross-validation requires prediction for the held-out case, which is impossible under the no-pooling model.\newline
The no-pooling model has the best raw fit to data, but after correcting for fitted parameters, the complete-pooling model has lowest estimated expected predictive error under the different measures.  In general, we would expect the hierarchical model to win, but in this particular case, setting $\tau=0$ (that is, the complete-pooling model) happens to give the best average predictive performance.
}\label{deviance.table}
\end{table}

For this model, the log predictive density is simply
$$
p(y|\theta) = \sum_{j=1}^J\log\left(\Nor(y_j|\theta_j,\sigma^2_j)\right)
= -\,\frac{1}{2}\sum_{j=1}^J\left(\log(2\pi\sigma_j^2)+ \frac{1}{\sigma_j^2}(y_j-\theta_j)^2\right).
$$
We shall go down the rows of Table \ref{deviance.table} to understand how the different information criteria work for each of these three models, then we discuss how these measures can be used to compare the models.

\paragraph{AIC.}  The log predictive density is higher---that is, a better fit---for the no pooling model.  This makes sense:  with no pooling, the maximum likelihood estimate is right at the data, whereas with complete pooling there is only one number to fit all 8 schools.  However, the ranking of the models changes after adjusting for the fitted parameters (8 for no pooling, 1 for complete pooling), and the expected log predictive density is estimated to be the best (that is, AIC is lowest) for complete pooling.  The last column of the table is blank for AIC, as this procedure is defined based on maximum likelihood estimation which is meaningless for the hierarchical model.

\paragraph{DIC.}  For the no-pooling and complete-pooling models with their flat priors, DIC gives results identical to AIC (except for possible simulation variability, which we have essentially eliminated here by using a large number of posterior simulation draws).  DIC for the hierarchical model gives something in between:  a direct fit to data (lpd) that is better than complete pooling but not as good as the (overfit) no pooling, and an effective number of parameters of 2.8, closer to 1 than to 8, which makes sense given that the estimated school effects are pooled almost all the way back to their common mean.  Adding in the correction for fitting, complete pooling wins, which makes sense given that in this case the data are consistent with zero between-group variance.

\paragraph{WAIC.}  This fully Bayesian measure gives results similar to DIC.  The fit to observed data is slightly worse for each model (that is, the numbers for lppd are slightly more negative than the corresponding values for lpd, higher up in the table), accounting for the fact that the posterior predictive density has a wider distribution and thus has lower density values at the mode, compared to the predictive density conditional on the point estimate. However, the correction for effective number of parameters is lower (for no pooling and the hierarchical model, $p_{\rm WAIC}$ is about half of $p_{\rm DIC}$), consistent with the theoretical behavior of WAIC when there is only a single data point per parameter, while for complete pooling, $p_{\rm WAIC}$ is only a bit less than 1, roughly consistent with what we would expect from a sample size of 8).  For all three models here, $p_{\rm WAIC}$ is much less than $p_{\rm DIC}$, with this difference arising from the fact that the lppd in WAIC is already accounting for much of the uncertainty arising from parameter estimation.

\paragraph{Cross-validation.}  For this example it is impossible to cross-validate the no-pooling model as it would require the impossible task of obtaining a prediction from a held-out school given the other seven. This illustrates on main difference to information criteria, which assume new prediction for these same schools and thus work also in no-pooling model.  For complete pooling and for the hierarchical model, we can perform leave-one-out cross-validation directly.  
In this model the local prediction of cross-validation is based only on the information coming from the other schools, while the local prediction in WAIC is based on the local observation as well as the information coming from the other schools. In both cases the prediction is for unknown future data, but the amount of information used is different and thus predictive performance estimates differ more when the hierarchical prior becomes more vague (with the difference going to infinity as the hierarchical prior becomes uninformative, to yield the no-pooling model). This example shows that it is important to consider which prediction task we are interested in and that it is not clear what $n$ means in asymptotic results that feature terms such as $o(n^{-1})$.


\paragraph{Comparing the three models.}
For this particular dataset, complete pooling wins the expected out-of-sample prediction competition.  Typically it is best to estimate the hierarchical variance but, in this case, $\tau=0$ is the best fit to the data, and this is reflected in the center column in Table \ref{deviance.table}, where the expected log predictive densities are higher than for no pooling or complete pooling.

That said, we still prefer the hierarchical model here, because we do not believe that $\tau$ is truly zero.  For example, the estimated effect in school A is 28 (with a standard error of 15) and the estimate in school C is $-3$ (with a standard error of 16).  This difference is not statistically significant and, indeed, the data are consistent with there being zero variation of effects between schools; nonetheless we would feel uncomfortable, for example, stating that the posterior probability is 0.5 that the effect in school C is larger than the effect in school A, given that data that show school A looking better.  It might, however, be preferable to use a more informative prior distribution on $\tau$, given that very large values are both substantively implausible and also contribute to some of the predictive uncertainty under this model.

In general, predictive accuracy measures are useful in parallel
with posterior predictive
checks to see if there are important patterns in the data that are
not captured by each model.  As with predictive checking, the log score can be computed
in different ways for a hierarchical model
depending on whether the parameters $\theta$ and
replications $y^{\rep}$ correspond to estimates and replications of new
data from the existing groups (as we have performed the calculations in the
above example) or new groups (additional schools from the $\Nor(\mu,\tau^2)$
distribution in the above example). 

\section{Discussion}

There are generally many options in setting up a model for any applied problem.
Our usual approach is to start with a simple model that uses only some of the available information---for example, not using some possible predictors in a regression, fitting a normal model to discrete data, or ignoring evidence
of unequal variances and fitting a simple equal-variance model.
Once we have successfully fitted a simple model, we can check its fit
to data and then alter or expand it as appropriate.

There are two typical scenarios in which models are compared.  First,
when a model is expanded, it is natural to compare the smaller to the
larger model and assess what has been gained by expanding the model (or,
conversely, if a model is simplified, to assess what was lost).  This
generalizes into the problem of comparing a set of nested models and judging
how much complexity is necessary to fit the data.

In comparing nested models, the larger model typically has the advantage of making more sense and fitting the data better but the disadvantage of being
more difficult to understand and compute.  The
key questions of model comparison
are typically:  (1) is the improvement in fit large enough to justify the additional difficulty in fitting, and (2) is the prior distribution on the
additional parameters reasonable?

The second scenario of model comparison is between two or more nonnested models---neither model generalizes the other. One might compare regressions that use different sets of predictors to fit the same data, for example, modeling political behavior using information based on past voting results or on demographics.  In these settings, we are typically not interested in {\em choosing} one of the models---it would be better, both in substantive and predictive terms, to construct a larger model that includes both as special cases, including both sets of predictors and also potential interactions in a larger regression, possibly with an informative prior distribution if needed to control the estimation of all the extra parameters.  However, it can be useful to {\em compare} the fit of the different models, to see how either set of predictors performs when considered alone.

In any case, when evaluating models in this way, it is important to adjust for overfitting, especially when comparing models that vary greatly in their complexity, hence the value of the methods discussed in this article.

\subsection{Evaluating predictive error comparisons}

When comparing models in their predictive accuracy, two issues arise, which might be called statistical and practical significance.  Lack of statistical significance arises from uncertainty in the estimates of comparative out-of-sample prediction accuracy and is ultimately associated with variation in individual prediction errors which manifests itself in averages for any finite dataset.  Some asymptotic theory suggests that the sampling variance of any estimate of average prediction error will be of order 1, so that, roughly speaking, differences of less than 1 could typically be attributed to chance, but according to Plummer (1996), this asymptotic result does not necessarily hold for nonnested models.
A practical estimate of related sampling uncertainty can be obtained by analyzing the variation in the expected log predictive densities $\widehat{\rm elppd}_i$ using parametric or nonparametric approaches (Vehtari and Lampinen, 2002).

Practical significance depends on the purposes to which a model will be used.  Sometimes it may be possible to use an application-specific scoring function that is so familiar for subject-matter experts that they can interpret the practical significance of differences. For example, epidemiologists are used to looking at differences in area under receiver operating characteristic curve (AUC) for classification and survival models. In settings without such conventional measures, it is not always clear how to interpret the magnitude of a difference in log predictive probability when comparing two models.  Is a difference of 2 important?  10?  100?  One way to understand such differences is to calibrate based on simpler models (McCulloch, 1989).  For example, consider two models for a survey of $n$ voters in an American election, with one model being completely empty (predicting $p=0.5$ for each voter to support either party) and the other correctly assigning probabilities of 0.4 and 0.6 (one way or another) to the voters.  Setting aside uncertainties involved in fitting, the expected log predictive probability is $\log(0.5)=-0.693$ per respondent for the first model and $0.6\log (0.6)+0.4\log(0.4)=-0.673$ per respondent for the second model.  The expected improvement in log predictive probability from fitting the better model is then $0.02n$.  So, for $n=1000$, this comes to an improvement of 20, but for $n=10$ the predictive improvement is only 2.  This would seem to accord with intuition:  going from 50/50 to 60/40 is a clear win in a large sample, but in a smaller predictive dataset the modeling benefit would be hard to see amid the noise.  

In our studies of public opinion and epidemiology, we have seen cases where a model that is larger and better (in the sense of giving more reasonable predictions) does not appear dominant in the predictive comparisons.  This can happen because the improvements are small on an absolute scale (for example, changing the predicted average response among a particular category of the population from 55\% Yes to 60\% Yes) and concentrated in only a few subsets of the population (those for which there is enough data so that a more complicated model yields noticeably different predictions).  Average out-of-sample prediction error can be a useful measure but it does not tell the whole story of model fit.

\subsection{Selection-induced bias}

Cross-validation and information criteria make a correction for using the data twice (in constructing the posterior and in model assessment) and obtain asymptotically unbiased estimates of predictive performance for a given model. However,  when these methods are used for model selection, the predictive performance estimate of {\em the selected model} is biased due to the selection process (see references in Vehtari and Ojanen, 2012).

If the number of compared models is small, the bias is small, but if the number of candidate models is very large (for example, the number of models grows exponentially as the number of observations $n$ grows, or the number of covariates $p \gg \ln(n)$ in covariate selection) a model selection procedure can strongly overfit the data. 
It is possible to estimate the selection-induced bias and obtain unbiased estimates, for example by using another level of cross-validation. This does not, however, prevent the model selection procedure from possibly overfitting to the observations and consequently selecting models with suboptimal predictive performance.  This is one reason we view cross-validation and information criteria as an approach for understanding fitted models rather than for choosing among them.

\subsection{Challenges and conclusions}

The current state of the art of measurement of predictive model fit remains unsatisfying.  Formulas such as AIC, DIC, and WAIC fail in various examples: AIC does not work in settings with strong prior information, DIC gives nonsensical results when the posterior distribution is not well summarized by its mean, and WAIC relies on a data partition that would cause difficulties with structured models such as for spatial or network data.  Cross-validation is appealing but can be computationally expensive and also is not always well defined in dependent data settings.

For these reasons, Bayesian statisticians do not always use predictive error comparisons in applied work, but we recognize that there are times when it can be useful to compare highly dissimilar models, and, for that purpose, predictive comparisons can make sense.  In addition, measures of effective numbers of parameters are appealing tools for understanding statistical procedures, especially when considering models such as splines and Gaussian processes that have complicated dependence structures and thus no obvious formulas to summarize model complexity.

Thus we see the value of the methods described here, for all their flaws.  Right now our preferred choice is cross-validation, with WAIC as a fast and computationally-convenient alternative.  WAIC is fully Bayesian (using the posterior distribution rather than a point estimate), gives reasonable results in the examples we have considered here, and has a more-or-less explicit connection to cross-validation, as can be seen its formulation based on pointwise predictive density (Watanabe, 2010, Vehtari and Ojanen, 2012).  A useful goal of future research would be a bridge between WAIC and cross-validation with much of the speed of the former and robustness of the latter.

\section*{References}

\noindent

\bibitem Aitkin, M. (2010).  {\em Statistical Inference:  An Integrated Bayesian/Likelihood Approach}.  London:  Chapman \& Hall.

\bibitem Akaike, H. (1973).  Information theory and an extension of the maximum likelihood principle.  In {\em Proceedings of the Second International Symposium on Information Theory}, ed.\ B. N. Petrov and F. Csaki, 267--281.  Budapest:  Akademiai Kiado.  Reprinted in {\em Breakthroughs in Statistics}, ed.\ S. Kotz, 610--624.  New York: Springer (1992).

\bibitem Ando, T., and Tsay, R. (2010).  Predictive likelihood for Bayesian model selection and averaging. {\em International Journal of Forecasting} {\bf 26}, 744--763.

\bibitem Bernardo, J. M. (1979). Expected Information as Expected Utility. {\em Annals of Statistics} {\bf 7}, 686--690.

\bibitem Bernardo, J. M., and Smith, A. F. M. (1994). {\em Bayesian Theory}. John Wiley \& Sons.

\bibitem Burman, P., Chow, E., and Nolan, D. (1994). A Cross-Validatory Method for Dependent Data. {\em Biometrika} {\bf 81}, 351--358.

\bibitem Burnham, K. P., and Anderson, D. R. (2002).  {\em Model Selection and Multimodel Inference:  A Practical Information Theoretic Approach}.  New York:  Springer.

\bibitem Celeux, G., Forbes, F., Robert, C. and Titterington, D. (2006). Deviance information criteria for missing data models. {\em Bayesian Analysis} {\bf 1}, 651–-706.

\bibitem DeGroot, M. H. (1970).  {\em Optimal Statistical Decisions}.  New York:  McGraw-Hill.

\bibitem Dempster, A .P. (1974). The direct use of likelihood for significance testing. Proceedings of Conference on Foundational Questions in Statistical Inference, Department of Theoretical Statistics: University of Aarhus, 335--352.

\bibitem Draper, D. (1999). Model uncertainty yes, discrete model averaging maybe. {\em Statistical Science} {\em 14}, 405--409.

\bibitem Efron, B., and Tibshirani, R. (1993).  {\em An Introduction to the Bootstrap}.  New York:  Chapman \& Hall.

\bibitem Geisser, S., and Eddy, W. (1979). A predictive approach to model selection. {\em Journal of the American Statistical Association} {\bf 74}, 153--160.

\bibitem Gelfand, A., and Dey, D. (1994). Bayesian model choice: asymptotics and exact calculations. {\em Journal of the Royal Statistical Society B} {\bf 56}, 501--514.

\bibitem Gelman, A., Carlin, J. B., Stern, H. S., and Rubin, D. B. (2003). {\em Bayesian Data Analysis}, second edition.  London:  CRC Press.

\bibitem Gelman, A., Meng, X. L., and Stern, H. S. (1996).  Posterior predictive assessment of model fitness via realized discrepancies
(with discussion).  {\em Statistica Sinica} {\bf 6}, 733--807.

\bibitem Gneiting, T. (2011). Making and evaluating point forecasts. {\em Journal of the American Statistical Association} {\bf 106}, 746--762.

\bibitem Gneiting, T., and Raftery, A. E. (2007). Strictly proper scoring rules, prediction, and estimation. {\em Journal of American Statistical Association} {\bf 102}, 359--378.

\bibitem Hibbs, D. (2008).  Implications of the `bread and peace' model for the 2008 U.S. presidential election.  {\em Public Choice} {\bf 137}, 1--10.

\bibitem Hoeting, J., Madigan, D., Raftery, A. E., and Volinsky, C. (1999).  Bayesian model averaging (with discussion). {\em Statistical Science} {\bf 14}, 382--417.

\bibitem Jones, H. E. and Spiegelhalter, D. J. (2012). Improved probabilistic prediction of healthcare performance indicators using bidirectional smoothing models. {\em Journal of the Royal Statistical Society A} {\bf 175}, 729--747.

\bibitem McCulloch, R. E. (1989). Local model influence. {\em Journal of the American Statistical Association} {\bf 84}, 473-–478.

\bibitem Plummer, M. (2008).  Penalized loss functions for Bayesian model comparison.  {\em Biostatistics} {\bf 9}, 523--539.

\bibitem Robert, C. P. (1996).  Intrinsic losses.  {\em Theory and Decision} {\bf 40}, 191--214.

\bibitem Rubin, D. B. (1981).  Estimation in parallel randomized experiments.
{\em Journal of Educational Statistics} {\bf 6}, 377--401.

\bibitem Rubin, D. B. (1984).  Bayesianly justifiable and relevant frequency
calculations for the applied statistician.  {\em Annals of Statistics}
{\bf 12}, 1151--1172.

\bibitem Schwarz, G. E. (1978). Estimating the dimension of a model. {\em Annals of Statistics} {\bf 6}, 461--464.

\bibitem Shibata, R. (1989).  Statistical aspects of model selection. In {\em From Data to Model}, ed.\ J. C. Willems, 215--240. Springer-Verlag.

\bibitem Spiegelhalter, D. J., Best, N. G., Carlin, B. P., and van der Linde, A. (2002).  Bayesian measures of model complexity and fit (with discussion). {\em Journal of the Royal Statistical Society B}.

\bibitem Spiegelhalter, D., Thomas, A., Best, N., Gilks, W., and Lunn, D.
(1994, 2003).
BUGS:  Bayesian inference using Gibbs sampling.  MRC Biostatistics Unit, Cambridge, England.\\
{\tt http://www.mrc-bsu.cam.ac.uk/bugs/}


\bibitem Stone, M. (1977). An asymptotic equivalence of choice of model cross-validation and Akaike's criterion. {\em Journal
of the Royal Statistical Society B} {\bf 36}, 44--47.

\bibitem van der Linde, A. (2005).  DIC in variable selection.  {\em Statistica Neerlandica} {\bf 1}, 45--56.

\bibitem Vehtari, A., and Lampinen, J. (2002).  Bayesian model assessment and comparison using cross-validation predictive  densities. {\em Neural Computation} {\bf 14}, 2439--2468.

\bibitem Vehtari, A., and Ojanen, J. (2012).  A survey of Bayesian predictive methods for model assessment, selection and comparison.  {\em Statistics Surveys} {\bf 6}, 142--228.

\bibitem Watanabe, S. (2009).  {\em Algebraic Geometry and Statistical Learning Theory}. Cambridge University Press.

\bibitem Watanabe, S. (2010).  Asymptotic equivalence of Bayes cross validation and widely applicable information criterion in singular learning theory. {\em Journal of Machine Learning Research} {\bf 11}, 3571--3594.

\bibitem Watanabe, S. (2013).  A Widely Applicable Bayesian Information Criterion. {\em Journal of Machine Learning Research} {\bf 14}, 867--897.

\end{document}